\documentclass[aps,pra,twocolumn,groupedaddress,showpacs,superscriptaddress,amssymb,amsmath]{revtex4-2}
\usepackage{natbib}
\usepackage[utf8]{inputenc}
\usepackage{graphicx}
\usepackage{tabularx}
\usepackage{xcolor}
\usepackage{amsmath}

\usepackage{comment}
\usepackage{dcolumn}
\usepackage{hyperref}
\definecolor{blue1}{HTML}{274057}
\hypersetup{
	colorlinks=true,
	linkcolor=blue,
	citecolor=blue,
    urlcolor=blue
}

\usepackage{bm}
\usepackage{epsf}
\usepackage{braket}
\usepackage{tensor}
\usepackage{soul}
\usepackage{mathrsfs}
\usepackage{mathtools}
\usepackage{multirow}

\usepackage{amsfonts}

\newcommand{\be}{\begin{equation}}
	\newcommand{\ee}{\end{equation}}
\newcommand{\bea}{\begin{eqnarray}}
	\newcommand{\eea}{\end{eqnarray}}

\def\la{\langle}

\def\ra{\rangle}

\makeatletter
\newsavebox{\@brx}
\newcommand{\llangle}[1][]{\savebox{\@brx}{\(\m@th{#1\langle}\)}%
	\mathopen{\copy\@brx\kern-0.5\wd\@brx\usebox{\@brx}}}
\newcommand{\rrangle}[1][]{\savebox{\@brx}{\(\m@th{#1\rangle}\)}%
	\mathclose{\copy\@brx\kern-0.5\wd\@brx\usebox{\@brx}}}
\makeatother

\begin{document}

\author{Katha Ganguly}
\email{katha.ganguly@students.iiserpune.ac.in}
\affiliation{Department of Physics,
		Indian Institute of Science Education and Research, Pune 411008, India}
        
\author{Bijay Kumar Agarwalla}
\email{bijay@iiserpune.ac.in}
\affiliation{Department of Physics,
		Indian Institute of Science Education and Research, Pune 411008, India}

\title{Measurement induced faster symmetry restoration in quantum trajectories}

\date{\today}

\begin{abstract}
Continuous measurement of quantum systems provides a standard route to quantum trajectories through the successive acquisition of information which further results in measurement back-action. In this work, we harness this back-action as a resource for global 
$U(1)$ symmetry restoration where continuous measurement is combined with a $U(1)$-preserving unitary evolution. Starting from a $U(1)$ symmetry-broken initial state, we simulate quantum trajectories generated by continuous measurements of both global and local observables. We show that under global monitoring, states containing superpositions of distant charge sectors restore symmetry faster than those involving nearby sectors. We establish the universality of this behavior across different measurement protocols. Finally, we demonstrate that local monitoring can further accelerate symmetry restoration for certain states that relax slowly under global monitoring. 
\end{abstract}

  \maketitle  

{\it Introduction.--}
Quantum dynamics of open systems has garnered significant attention in recent years ~\cite{Shiwu1998,Yan2002,PeterZoller2010,Franco2011,IROct2015,Kapral_2015,HPBApr2016,Daniel2017,Adolfo2017,Plenio2018,Zhong2019,Dries2020,Schiro_2024,Marco2024,Aurelia2025,PN2025,Mariia2025,dutta2025,Sciro2025,Ganguly_2025}, as it encompasses a broad spectrum of research areas such as quantum transport \cite{AMJFeb1982,LBMar2015,SGJul2017,JSBJan2018,Caspel2018,Archak2018,Zhang2020,Prosen2023,DSB2024,YPW2024}, relaxation dynamics~\cite{ThomasBarthel2013,Torres-Herrera_2014,Znidaric2015,Igor2021,Igor2022,Chatterjee2023,Wang2023,Adam2024,beato2025,Dante2025}, etc. Beyond the traditional open quantum system framework based on quantum master equations~\cite{BPOQS,carmichael1998,L1976,VG1976}, recent work has increasingly focused on stochastic quantum trajectories~\cite{wiseman2010,jacobs2014,MBPJan1998} which appear as unravelings of the quantum master equation~\cite{NGisin_1992,Molmer2008,carmichael2009,GISIN1992315}. While the quantum master equation gives an ensemble averaged description of open system evolution, stochastic quantum trajectories on the other hand offer deeper insight by resolving the dynamics of individual ensemble realizations. One of the common ways to generate quantum trajectories is via continuous monitoring of the quantum system along with the unitary evolution~\cite{Jacobs01092006,ToddBrun2000,Tamir2013,mensky2017,Landi2024,YanBin2026}. 
Recent studies of continuous measurement and quantum trajectories have uncovered a variety of phenomena—including measurement-induced entanglement transitions~\cite{Nahum_2019,Romito2019,Sebastian2021,Khemani-Huse2021,Saito2022,Muller2022,Russomano_2022,Fazio2025}, spontaneous symmetry breaking~\cite{Adolfo_2019}, charge-sharpening transitions~\cite{Sarang_2022}, learnability transitions~\cite{Andrew2022,Khemani2024,vasseur2025}, purification transition~\cite{Huse2020,anzai2025}, Zeno phase transitions~\cite{Phillip2025}, etc.—that remain hidden at the level of ensemble-averaged dynamics. Advances in quantum technological platforms~\cite{Kraus2008,Bloch2012,Blatt2012,PeterZoller2018,Blatt2019,WU2021213,Schäfer2020,Schäfer2020} have further propelled  the study of quantum dynamics by realizing individual stochastic trajectories with high controllability~\cite{Rainer1986,Vijay2011,Kater2013}. 

Alongside quantum trajectories, understanding relaxation dynamics toward steady states has also attracted significant interest for a variety of reasons. Characterization of relaxation dynamics plays a key role in identifying universality classes in quantum transport \cite{Ljubotina2017,Sarang2022}. Furthermore, accelerated relaxation is relevant to processes that involves resetting \cite{Ranjan2024,solanki2025,Bao2025}. Recent works have explored faster relaxation of farther from equilibrium quantum states which is the interesting quantum Mpemba effect in open quantum systems~\cite{Andrea2024,longhi2025mpemba,bagui2025,liu2025,Zhang_Mpemba2025,caldas2025,Archak2025,ulcakar2025,bagui_chatterjee2025}. Most of these works are based on ensemble averaged description of open quantum systems.
Several studies have also investigated relaxation and dynamical symmetry restoration in local subsystems of isolated quantum systems~\cite{Liu2024,Yamashika2024,Filiberto2024,Xhek2025,di2025,Yu2025,Tanmay2025,Heng2025}. However a thorough understanding of relaxation dynamics in quantum trajectories is still lacking. 
\begin{figure}
    \centering
 \includegraphics[width=\linewidth]{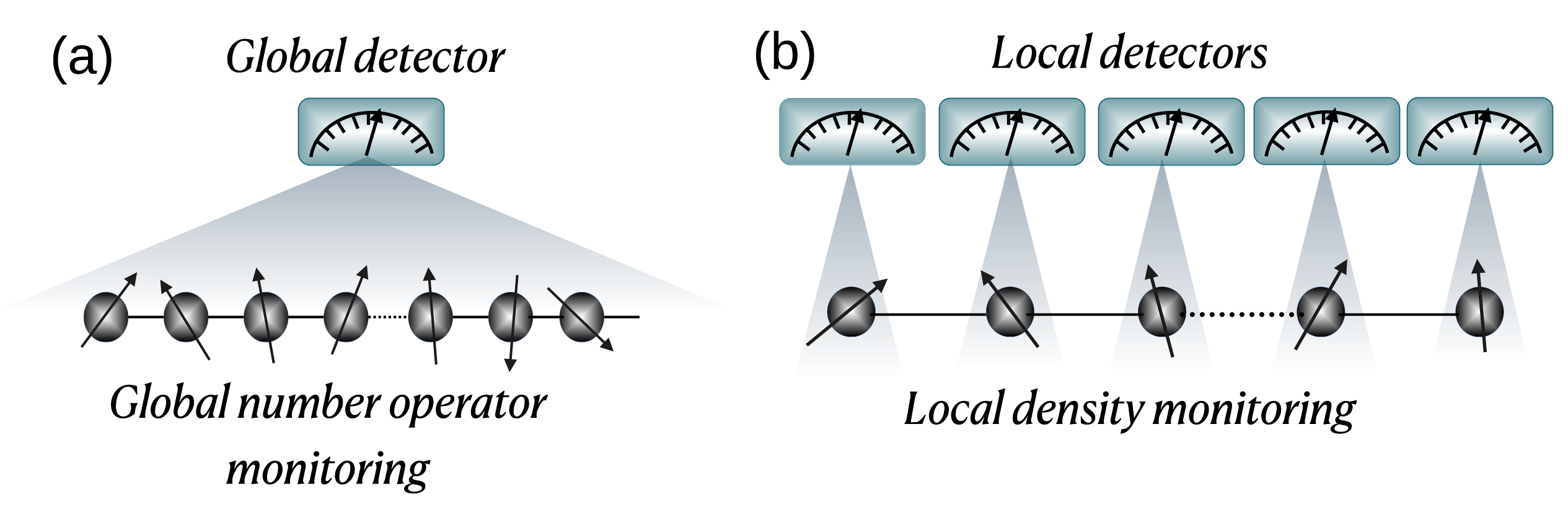}
 \includegraphics[width=\linewidth]{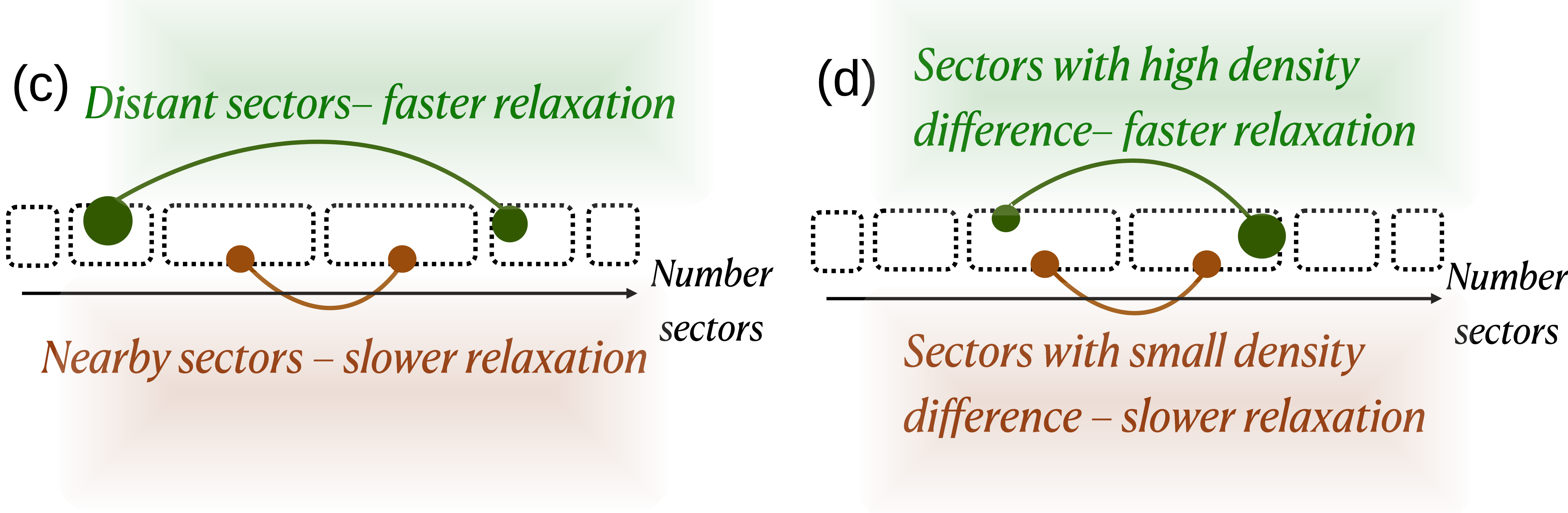}
    \caption{Schematic of the setup:  one dimensional spin-1/2 lattice of length $L$, represented by XX Hamiltonian respecting $U(1)$ symmetry, is subjected to continuous monitoring of the (a) global operator $\hat{N}=\hat{S}_z+L/2$, which describes the total number of up spins in the lattice, and (b) local number operator/density $\hat{n}_i=\hat{S}_z^i+1/2$ at every lattice site. Starting from a broken $U(1)$ state, under such global or local continuous monitoring, $U(1)$ symmetry can be dynamically restored. We have schematically represented our findings in (c) \& (d) for global and local monitoring, respectively. Under global monitoring, a symmetry broken state that contains distant number sectors restores symmetry faster than a state involving nearby number sectors. However in presence of local monitoring, the relaxation dynamics depends on the density profile differences across different number sectors. A state containing number sectors that have higher density difference relaxes faster than a state involving sectors with smaller density difference.}
    \label{fig:schematic}
\end{figure}

In this Letter, we investigate relaxation dynamics in quantum trajectories associated with the dynamical restoration of the global $U(1)$ symmetry. Initializing the system in a $U(1)$ symmetry-broken state, we first study the subsequent time evolution under the combined action of a $U(1)$-conserving Hamiltonian and continuous monitoring of the corresponding global conserved charge. At the level of individual quantum trajectories, the initially symmetry-broken state relaxes toward a globally $U(1)$-symmetric state through measurement back-action, which progressively drives the system toward an eigenstate of the conserved charge. We show that the timescale of this symmetry restoration depends on the separation between charge sectors in the initial superposition: states involving distant charge sectors restore symmetry faster than those involving nearby sectors. We establish the universality of this behavior across different continuous monitoring protocols. Finally, we extend our analysis to continuous monitoring of local observables and identify key qualitative differences between symmetry restoration under global and local measurements. 

\vspace{0.2cm}
\textit{Setup.--}
We consider a generic one-dimensional spin-1/2 lattice of size $L$ described by a $U(1)$ conserving Hamiltonian $H_S$ that conserves the total magnetization $\hat{S}_z=\sum_{i=1}^L \hat{S}_{z}^i$, or equivalently the total number of up spins $\hat{N}=\hat{S}_z+L/2$ in the lattice (henceforth referred to as number operator). Under a pure unitary evolution generated by the Hamiltonian $H_S$, an initial state that breaks the $U(1)$ symmetry can exhibit symmetry restoration only at the level of local subsystems, but not in the global state of the system \cite{Filiberto2024,Ares2025,Tanmay2025}. One possible way to restore the symmetry globally through the dynamics would be to combine the unitary evolution with continuous measurements of either the total number operator $\hat{N}$ or the local number operator $\hat{n}_i=\hat{S}_z^i+1/2$ throughout the lattice, as illustrated in Fig.~\ref{fig:schematic}(a) and~\ref{fig:schematic}(b), respectively. Note that we purposefully avoid the continuous monitoring of total magnetization $\hat{S}_z$ as it is not always efficient to restore $U(1)$ symmetry globally [see End Matter].
The system is prepared in a symmetry broken initial state, $|\Psi_0\ra=\sum_{n=0}^{L}c^0_n|n\rangle$, where $|n\ra$ is an eigenstate of the number operator $\hat{N}$ with eigenvalue $n$ that represents $n$ number of up spins in the lattice, and $c_n$ is the probability amplitude corresponding to $|n\ra$. Note that, the state $|n\ra$ can, in general, represent any superposition of different configurations with fixed number of up and down spins.
In the present analysis, we have considered the XX spin-chain with Hamiltonian $H_S=\sum_{i=1}^{L-1}J\big[\hat{S}_x^i\hat{S}_x^{i+1}+\hat{S}_y^{i}\hat{S}_y^{i+1}\big]$ with $J$ being the hopping strength. However, our results hold for other $U(1)$ conserving Hamiltonians as well. In the following, we first discuss the symmetry restoration under the global monitoring of the operator $\hat{N}$ by employing two prototypical measurement protocols-- Quantum Jump (QJ) protocol \cite{Zoller1992,Molmer_93,MBPJan1998} and Quantum State Diffusion (QSD) protocol~\cite{NGisin_1992,GISIN1992315,Jacobs01092006}. 
Furthermore, to corroborate the universality of our results across different protocols, we devise a general continuous positive operator valued measurement (POVM) and discuss the symmetry restoration. Towards the end, we extend our discussion to local continuous measurements and showcase the intricate differences between the global and local measurements. 

\vspace{0.2cm}
\textit{Quantum Jump (QJ) protocol.--}
QJ process describes a particular type of continuous monitoring, where the information about the system is extracted either by a direct jump from one state to the other or by a smooth no jump evolution by an effective non-Hermitian Hamiltonian at each instant of time. Such protocol is experimentally motivated by direct photo-detection processes~\cite{wiseman2010}.
We consider the continuous monitoring of the total number of up spins $\hat{N}$ by the QJ protocol. Note that, instead, if one measures the total magnetization $\hat{S}_z$ of the lattice, surprisingly, the QJ process is inefficient to restore the $U(1)$ symmetry [see End Matter]. The stochastic Schr\"odinger equation (SSE) describing the evolution of the system under the QJ protocol is given by \cite{Landi2024},
\begin{align}
   d|\Psi_t\ra = \!\!\Bigg[\!-i\hat{H}_S \, dt - &\frac{\gamma\, dt}{2}\big(\hat{N}^2-\la \hat{N}^2\ra_t\big)\nonumber\\&+\Bigg(\frac{\hat{N}}{\sqrt{\la \hat{N}^2\ra_t}}-\mathbb  I\Bigg)d\chi_t\Bigg]|\Psi_t\ra, \label{eq:SSE_QJ}
\end{align}
where $d\chi_t$ is a Poisson noise which is $1$ with probability $P(d\chi_t\!=\!1)=\gamma \, dt \, \la \hat{N}^2\ra_t$ and $0$ with probability $1-P(d\chi_t\!=\!1)$. Here $\mathbb{I}$ is the identity operator with dimension $2^{L}\times 2^L$ and $\gamma$ is the strength of monitoring. Note that the no jump trajectory can be constructed by considering only the first two terms in Eq.~\eqref{eq:SSE_QJ} which mimicks a deterministic evolution by a non-Hermitian Hamiltonian $\hat{H}_S-i\gamma\hat{N}^2/2$ with normalization. A jump corresponds to the activation of the last term in Eq.~\eqref{eq:SSE_QJ} in which the jump operator $\hat{N}$ acts on the state. Starting from a symmetry-broken initial state, that contains a superposition of different eigenstates of $\hat{N}$ (which constructs different number sectors), the dynamics generated by the SSE in Eq.~\eqref{eq:SSE_QJ} does not introduce any further new eigenstates into the superposition; instead, it only modifies the probability amplitudes of the existing components i.e., $|\Psi_t\ra=\sum_{n=0}^{L}c_n^t|n\ra$, through unitary evolution and measurement-induced back-action. Here $c_n^t$ is the probability amplitude at time $t$ corresponding to the eigenstate $|n\ra$ and following Eq.~\eqref{eq:SSE_QJ}, the evolution of $|c_n^t|^2$ is given as,
\begin{align}
d|c_n^t|^2=|c^t_n|^2\Big[\frac{d\chi_t}{\la \hat{N}^2\ra_t}-\gamma dt\Big]\sum_{m\neq n}|c_m^t|^2\big(n^2-m^2\big).\label{eq:dc_QJ}
\end{align}
Here, $c_n^t$ represents a collective probability amplitude, encompassing superpositions of different configurations of up and down spins within the $n$-th number sector. Consequently, it can acquire only a global phase by the action of Hamiltonian $\hat{H}_S$, which does not affect the probability $|c_n^t|^2$.
Eq.~\eqref{eq:dc_QJ} further shows that the evolution of the probabilities $|c^t_n|^2$ is controlled by the separations between the number sectors, quantified by the differences $n^2 - m^2$. The appearance of such terms are purely due to measurement back-action, which generates different time-scales in the dynamics. Each time-scale is inversely proportional to one of the differences $n^2-m^2$.
The smallest such difference sets the longest timescale at which $d|c_n^t|^2 \approx 0$ which is the time-scale of symmetry restoration. Although the symmetry restoration timescale depends on smallest value of $n^2-m^2$, due to the presence of noise $d\chi_t$, this can be different in different quantum trajectories~\footnote{\label{fn:pathology}There might be some pathological case in which one post-selects a quantum trajectory which corresponds to $\gamma dt\la\hat{N}^2\ra_t\approx d\chi_t$ for all time $t$. In such case, in spite of having superposition of distant sectors, the rate of symmetry restoration is still very small. However, for large system sizes, such trajectories are unlikely to appear. Similar cases exist for QSD as well which corresponds to $d\xi_t\approx0$ at each time $t$.}. 
Using Eq.~\eqref{eq:SSE_QJ}, we also obtain the evolution of the average number of up or down spins in the system, i.e., $\la \hat{N}\ra_t$ in each trajectory, which is given by,
\begin{equation}
    d\la \hat{N}\ra_t=\Big[\frac{d\chi_t}{\la \hat{N}^{2}\ra_t}-\gamma dt\Big]\Big(\la \hat{N}^3\ra_t-\la \hat{N}\ra_t\la \hat{N}^2\ra_t\Big). \label{eq:S_z_evol_QJ}
\end{equation}
The Hamiltonian respecting $U(1)$ symmetry does not contribute to the evolution of $\la \hat{N}\ra_t$ in Eq.~\eqref{eq:S_z_evol_QJ}.
\begin{figure}
    \centering
    \includegraphics[width=0.5\linewidth]{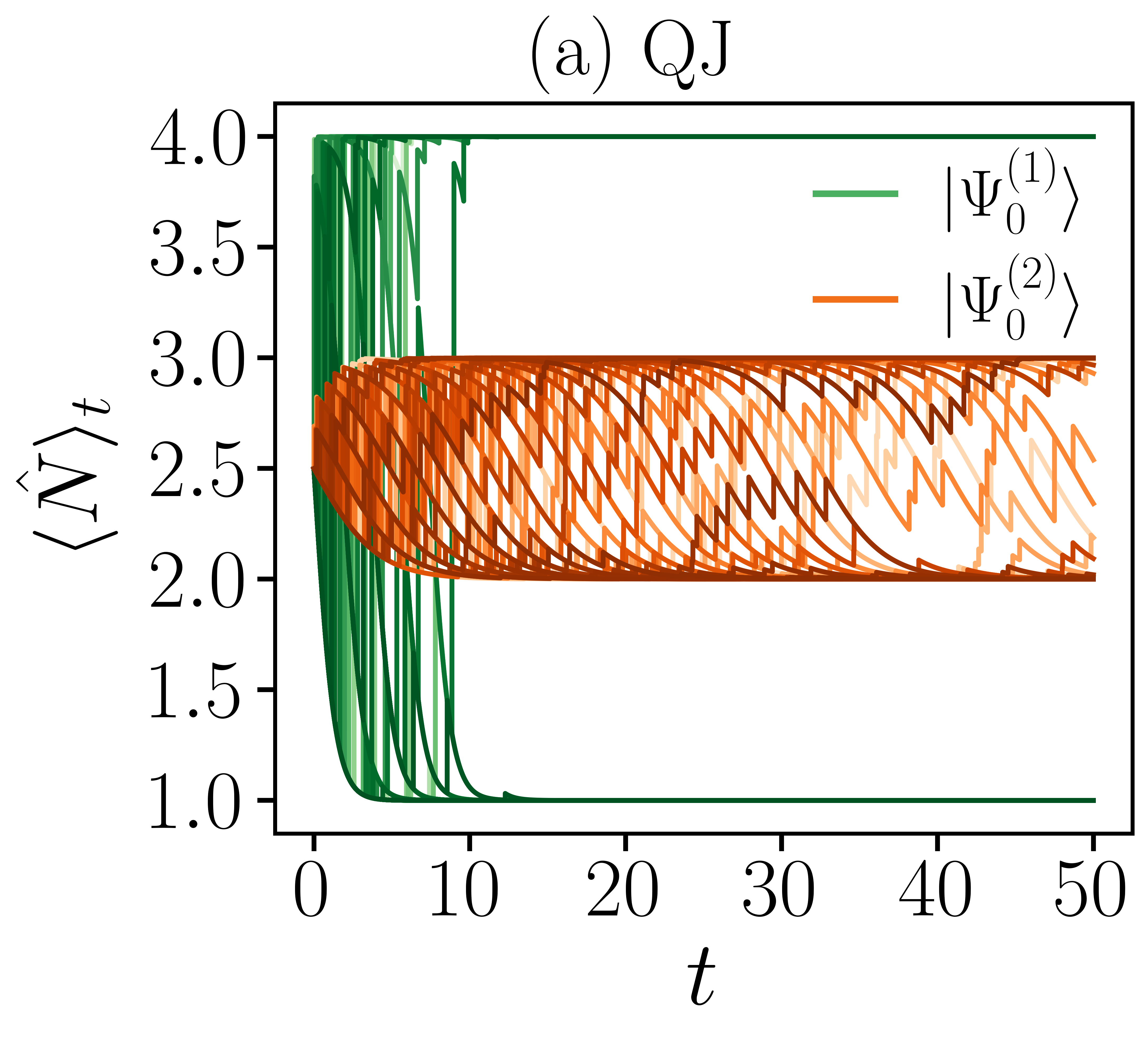}%
    \includegraphics[width=0.5\linewidth]{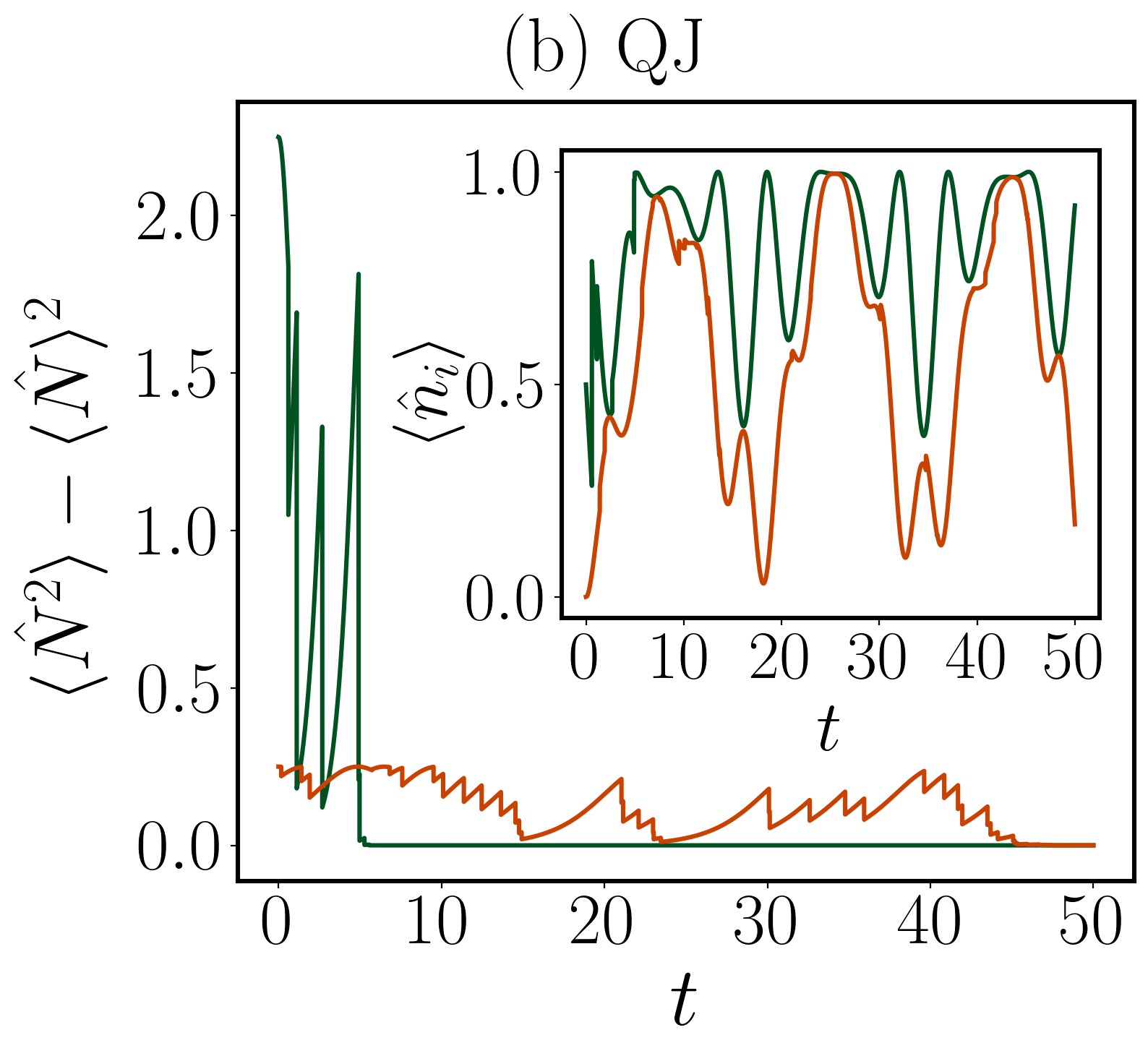}
    \caption{Plot for the quantum jump (QJ) protocol: (a) The dynamics of the total number of up spins in the lattice i.e, $\la \hat{N}\ra=\la\hat{S}_z\ra+L/2$ is plotted for $100$ different trajectories for both initial states $|\Psi_0^{(1)}\ra$ (green solid) and $|\Psi_0^{(2)}\ra$ (brown solid), as mentioned in the main text. Here $\la \hat{N}\ra_{\rm ss}$ is the number of up spins in the lattice after the symmetry restoration which is either $n$ or $L-n$ in a trajectory.
    (b) The dynamics of fluctuation in $\hat{N}$ is plotted for the same initial states. The initial state $|\Psi_0^{(1)}\ra$, which has larger initial fluctuation shows faster symmetry restoration than the state $|\Psi_0^{(2)}\ra$  which has smaller number fluctuation. The inset in (b) represents the dynamics of local number  operator $\la \hat{n}_i\ra= \hat{S}_z^i + 1/2$ at site $i=3$ which shows unitary dynamics with oscillations after the symmetry restoration. For the numerics, we have considered $L=5$, $\gamma=0.1$, $dt=0.01$. The unitary evolution is governed by the XX Hamiltonian with hopping strength $J=1$. The central results for $\la \hat{N}\ra$ and its fluctuations are however independent of the specific form of the lattice Hamiltonian.}
    \label{fig:QJ}
\end{figure}
Note that the dynamics of $\la \hat{N}\ra_t$ is governed by the factor $\la \hat{N}^3\ra_t-\la \hat{N}\ra_t\la \hat{N}^2\ra_t$ which can be simplified to $\sum_{n}\sum_{m\neq n}|c_n^t|^2|c_m^t|^2n^2(n-m)$, which also depends on the separations between different number sectors present in the state $|\Psi_t\ra$.
The smallest separation decides the time-scale at which $d\la\hat{N}\ra_t\approx0$ and at this time-scale, the state gets confined to any one of the number sectors  of the initial state. Needless to mention, although the Hamiltonian $H_S$ commutes with the total number operator $\hat{N}$, an eigenstate of $\hat{N}$ i.e., the $U(1)$ symmetry restored state is not necessarily an eigenstate of $\hat{H}$ due to degeneracies in the eigenvalues of $\hat{N}$. As a result, after symmetry restoration, the system may evolve under a pure unitary dynamics governed by the lattice Hamiltonian $H_S$. 

In Fig.~\ref{fig:QJ}, we present numerical results of symmetry restoration following quantum jump trajectories by performing exact diagonalization. We consider two initial states of the form $|\Psi_0\ra=\big(|n\ra+|L\!-\!n\ra\big)/\sqrt{2}$, where $L$ denotes the system size. The first one corresponds to $n=1$ and is denoted by $|\Psi_0^{(1)}\ra$, while the second corresponds to $n=(L\!-\!1)/2$ and is denoted by $|\Psi_0^{(2)}\ra$. In Fig.~\ref{fig:QJ}(a), we observe that all the trajectories originating from $|\Psi_0^{(1)}\ra$ restore the symmetry {\it faster} than those generated from $|\Psi_0^{(2)}\ra$. Here we plot the time evolution of the expectation value $\langle \hat{N} \rangle_t$ for $100$ different quantum trajectories. In each trajectory, $\langle \hat{N} \rangle_t$ eventually settles to one of the eigenvalues of $\hat{N}$. The dynamics for both the initial states has only one time-scale which is inversely proportional to $(n-L/2)^2$. Hence, the state $|\Psi_0^{(1)}\ra$ has exponentially faster time-scale of symmetry restoration than $|\Psi_0^{(2)}\ra$. In Fig.~\ref{fig:QJ}(b), we plot the dynamics of number fluctuations along individual quantum trajectories, which eventually decay to zero, signaling the symmetry restoration. The state with larger initial fluctuations, $|\Psi_0^{(1)}\ra$ restores the symmetry faster than $|\Psi_0^{(2)}\ra$ which has smaller initial fluctuations. However, it is important to note that this accelerated, measurement-induced symmetry restoration is not determined by the magnitude of the initial fluctuations, but rather by the separation between the number sectors present in the initial superposition, i.e., $(2n-L)$ in our case. In fact, there are states with higher number fluctuations that restore the symmetry slower than a state with smaller number fluctuation [see supplementary material]. This happens when both the two states involves equidistant number sectors in the superposition. In the inset of Fig.~\ref{fig:QJ}(b), we have plotted the local number operator $\hat{n}_i=\hat{S}_z+1/2$ with time $t$ which shows that beyond the symmetry restoration time-scale, $\la \hat{n}_i\ra_t$ shows oscillatory dynamics, purely due to the effect of lattice Hamiltonian $\hat{H}_S$. In this time-scale, the global measurement of $\hat{N}$ does not play any further role.
\begin{figure}
    \centering
    \includegraphics[width=0.5\linewidth]{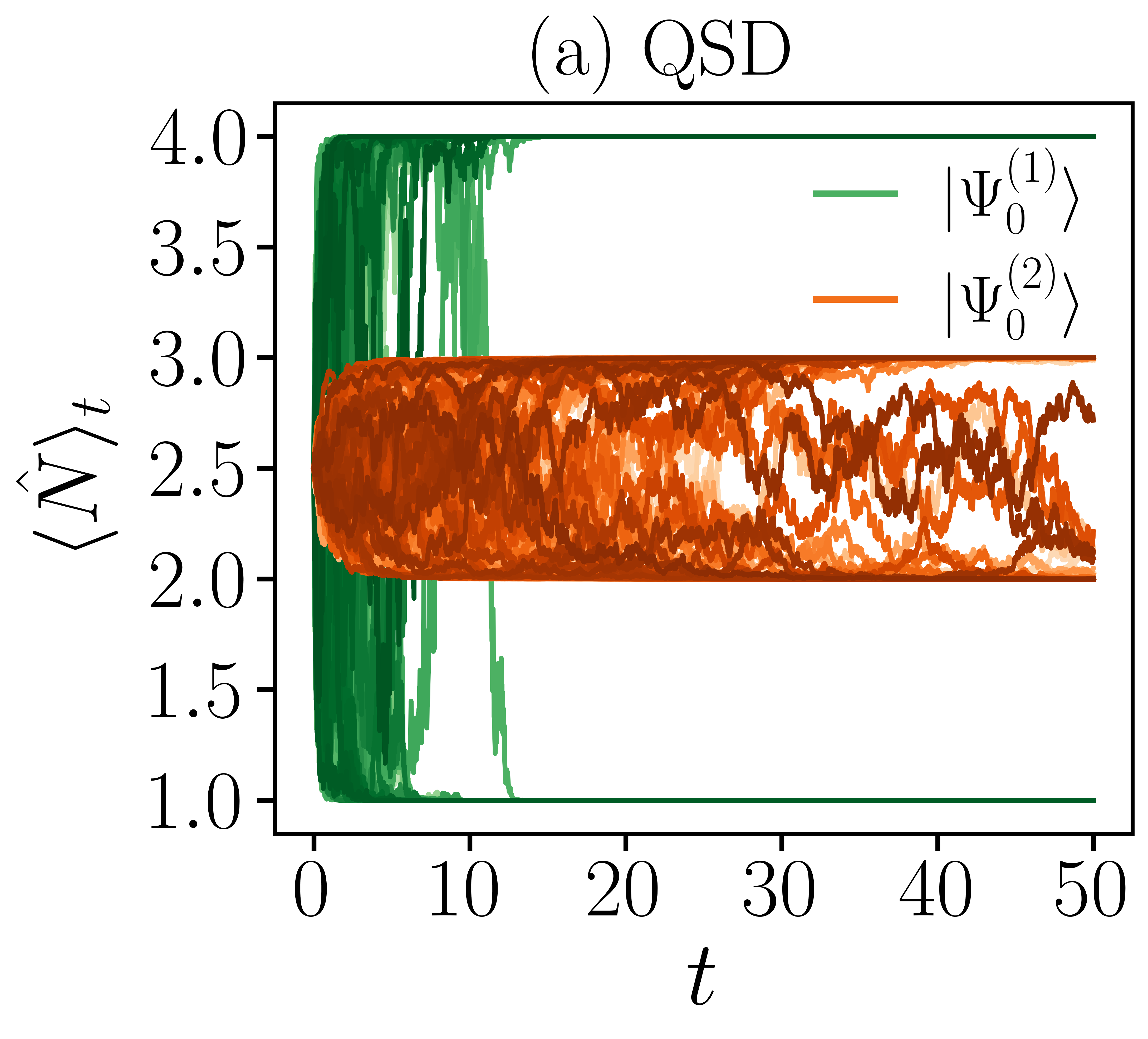}%
    \includegraphics[width=0.5\linewidth]{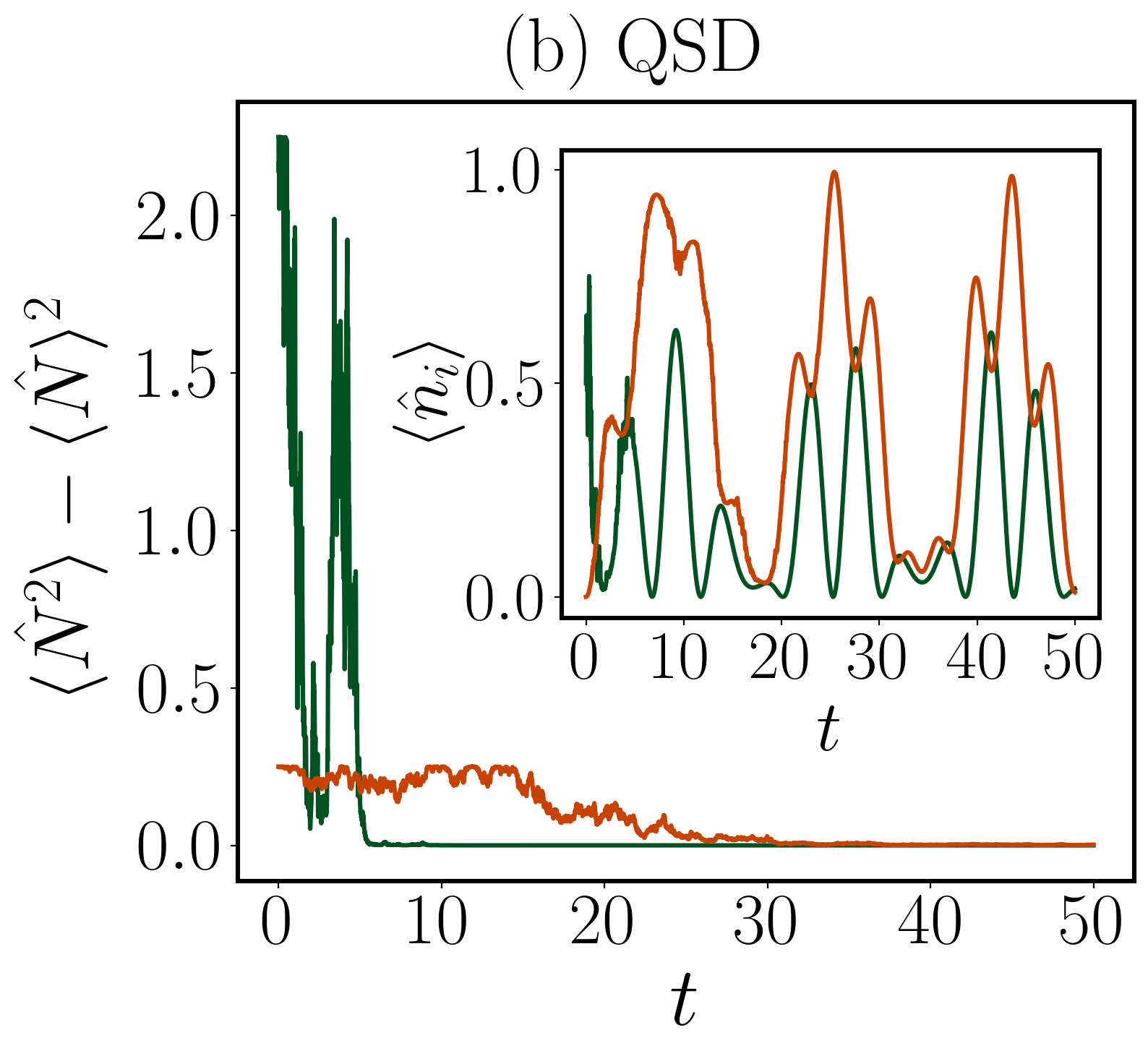}
    \caption{Plot for the quantum state diffusion (QSD) protocol: (a) The dynamics of $\la \hat{N}\ra=\la\hat{S}_z\ra+L/2$ is plotted for $100$ different trajectories for the same two initial states as Fig.~\ref{fig:QJ}. (b) The fluctuation in $\hat{N}$ is plotted with time $t$ for those initial states. The initial state $|\Psi_0^{(1)}\ra$, which has larger initial fluctuation shows faster symmetry restoration than the state $|\Psi_0^{(2)}\ra$  which has smaller number fluctuation. The inset in (b) represents the behaviour of local magnetization $\la \hat{n}_i\ra=\hat{S}_z^i+1/2$ at site $i=3$ with time which shows unitary dynamics with oscillations after the symmetry restoration. The other parameters are the same as in Fig.~\eqref{fig:QJ}.}
    \label{fig:QSD}
\end{figure}

\vspace{0.2cm}
\textit{Quantum State Diffusion (QSD) protocol.--}
We next discuss global $U(1)$ symmetry restoration under the quantum state diffusion (QSD) protocol for global monitoring, which corresponds to another type of continuous weak measurement of the system~\cite{wiseman2010,Jacobs01092006}. In QSD, the acquisition of information at each time instant is infinitesimal which produces a tiny back-action at each instant of time. This protocol is experimentally motivated by homodyne and heterodyne detection processes~\cite{Yuen_83,Warszawski_2002,wiseman2010}. The SSE that governs the evolution of trajectories under QSD is given by,
\begin{equation}
    d|\Psi_t\rangle\! =\!\Big[\!\! -i\hat{H}_S dt\!-\!\frac{\gamma dt}{2}(\hat{N}\!-\!\langle\hat{N}\rangle_t)^2+(\hat{N}\!-\!\langle\hat{N}\rangle_t)d\xi_t\Big]|\Psi_t\ra, \label{eq:SSE_QSD}
\end{equation}
where $d\xi_t$ is Wiener increment with the statistics; $\overline{d\xi_t}=0$, $\overline{d\xi_t^2}=\gamma dt$. Similar to the QJ case, here also, the state at any instant of time $t$ can be written as $|\Psi_t\ra=\sum_{n=0}^{L}c_n^t|n\ra$ and
following Eq.~\eqref{eq:SSE_QSD}, the evolution of the probability $|c_n^t|^2$ can be obtained as,
\begin{equation}
    d|c_n^t|^2 = 2 \,|c_n^t|^2 d\xi_t\,\sum_{m\neq n}|c_m^t|^2(n-m).\label{eq:probability_qsd}
\end{equation}
We also obtain the evolution of the average number of particles $\la \hat{N}\ra_t$, using Eq.~\eqref{eq:SSE_QSD} as,
\begin{align}
    d\la \hat{N}\ra_t=2\, d\xi_t \, \big[ \la \hat{N}^2\ra_{t}-\la\hat{N}\ra_t^2\big].\label{eq:Ntot}
\end{align}
Note that the evolution of $\la \hat{N}\ra_t$ is governed by the total number fluctuation which can be expressed as $\sum_{n}\sum_{m\neq n}|c_n^t|^2|c_m^t|^2(n-m)^2$. Interestingly, the time-scale of symmetry restoration is governed by the smallest value of the difference $(n-m)^2$, in the QSD case as well. Similar to the QJ protocol, we have simulated quantum trajectories for QSD following Eq.~\eqref{eq:SSE_QSD} for lattice size $L=5$, generated from the initial state of the form $|\Psi_0\ra=\big(|n\ra+|L-n\ra\big)/\sqrt{2}$. As mentioned before, this initial state  has only one time scale, inversely proportional to $(n-L/2)^2$. As the initial state $|\Psi_0^{(1)}\ra$ is associated with a very fast time scale, the trajectories generated from this state restores the symmetry faster than the trajectories generated from $|\Psi_0^{(2)}\ra$. This can be clearly observed in Fig.~\ref{fig:QSD}(a), where we have plotted $\la\hat{N}\ra_t$. 
In Fig.~\ref{fig:QSD}(b), we have plotted the number fluctuations for these two states which shows a crossing that further confirms the faster symmetry restoration of $|\Psi_0^{(1)}\ra$ compared to $|\Psi_0^{(2)}\ra$. The inset in Fig.~\ref{fig:QSD}(b) represents the unitary oscillations of the local number operator $\la \hat{n}_i\ra$ for $i=3$ with time $t$. One can further generalize Eq.~\eqref{eq:Ntot} to higher order cumulants, and for QSD protocol, interestingly the evolution of the any $k$-th cumulant is governed by the $(k+1)$-th cumulant at each trajectories [see supplementary Material]. It is important to mention that, both QJ and QSD protocol, at the ensemble level produces the same Gorini - Kossakowski - Sudarshan - Lindblad (GKSL) quantum master equation~\cite{L1976,VG1976,BPOQS}, given as, $\partial\rho/\partial t =-i[\hat{H},\rho]+\gamma\,\big[\hat{N}\rho  \hat{N}-\frac{1}{2}\{\hat{N}^2,\rho\}\big]$ which does not reveal any relaxation signature in $\la \hat{N}\ra_t$ even starting with a $U(1)$ broken state [see Supplementary material].

\vspace{0.2cm}
\textit{General conditional system-ancilla dynamics.--} We now generalize our results of global monitoring of the operator $\hat{N}$ by conditioning the system–bath dynamics to emulate a general continuous positive-operator-valued measurement (POVM). We model the bath as an infinite sequence of identical ancillas, each initialized in the state $|\Psi_A\ra$. The system interacts sequentially with the ancillas, one at a time, via the unitary operator $U=\exp{(-i\epsilon\hat{N}\otimes\hat{A})}$ where $\hat{A}$ is a Hermitian ancilla operator and $\epsilon$ controls the system-ancilla coupling strength. To extract information about the system, a projective measurement described by the projectors $\{|\mu_i\ra\la\mu_i|,i=1,\dots,d_{A}\}$ is performed on the ancilla, where $d_A$ denotes the dimension of the ancilla Hilbert space. This measurement induces a corresponding measurement back-action on the system. After the measurement, the system and ancilla are decoupled, thereby completing one step of the conditional system–ancilla dynamics. In the subsequent step, a fresh ancilla is coupled to the system and the same protocol is repeated.
The normalized state of the system conditioned by the outcome $\mu_i$ of the projective measurement on an ancilla after the first step is,
\begin{align}
    |\Psi_S(\mu_i)\ra=\sum_n c^1_n|n\ra=\frac{\sum_{n}c^0_nd_{\mu_i}(n)|n\ra}{\sqrt{\sum_{m}|c^0_m|^2|d_{\mu_i}(m)|^2}}\,. \label{eq:P-M_state}
\end{align}
Here $d_{\mu_i}(n)=\la\mu_i|e^{-i\epsilon n\hat{A}}|\Psi_A\ra$ and it modifies the probability amplitude corresponding to $|n\ra$. Therefore, the change in the probability $|c^0_n|^2$ under this protocol is given as,
\begin{equation}
    |c^1_n|^2\!\!-|c^0_n|^2=\frac{\sum_{m\neq n} |c^0_n|^2|c^0_m|^2\Big(|d_{\mu_i}(n)|^2-|d_{\mu_i}(m)|^2\Big)}{\sum_m |c_m^0|^2|d_{\mu_i}(m)|^2}.
\end{equation}
Note that if, for most measurement outcomes $\mu_i$, the quantities $|d_{\mu_i}(n)|^2$ are equal for any two eigenvalues $n$ and $n'$, then the continuous POVM cannot distinguish between the two eigenstates $|n\ra$ and $|n'\ra$. This follows from the fact that the relative change in the occupation probability of the eigenstate $|n\ra$, given by $ (|c_n^1|^2-|c_n^0|^2)/|c_n^0|^2$, is identical to that of $|n'\ra$. Consequently, the probabilities $|c_n^t|^2$ and $|c_{n'}^t|^2$ evolve identically: they either both decay to zero or both grow and saturate at finite values. Such a situation arises in the quantum jump (QJ) protocol under the measurement of the total magnetization$\hat{S}_z$. For both the jump and no-jump evolution, one finds $|d_\mu(-\mathcal{S}_z)|^2=|d_\mu(\mathcal{S}_z)|^2$ where $\pm \mathcal{S}_z$ are the eigenvalues of $\hat{S}_z$ and as a consequence symmetry is not always restored [see End Matter].  Another example is $d_{\mu_i}(n)=e^{-in\phi(\mu_i)}$, which arises when the projective measurement on the ancilla is performed in the eigenbasis of the coupling operator $\hat{A}$. Therefore, within this measurement scheme, symmetry restoration is not possible in any trajectories.

Here, we discard these possibilities and restrict our attention to POVMs that can distinguish all eigenstates of the measured observable. Notably, within this general scenario, the time scale of symmetry restoration is inversely proportional to the smallest value of $|d_{\mu}(n)|^2-|d_{\mu}(m)|^2$. For the quantum state diffusion (QSD) and quantum jump (QJ) protocols, this quantity always reduces to a simple function of $n-m$. For this general scenario, we focus 
on the limit $\epsilon \ll 1/n$ ($n$ is the largest eigenvalue of $\hat{N}$), the function $d_{\mu}(n)$ can be expanded as $d_{\mu}(n)=\la\mu|\Psi_A\ra-i\,\epsilon\, n\,\la\mu|\hat{A|\Psi_A\ra}$. Retaining terms up to $O(\epsilon)$, the difference $|d_{\mu_i}(n)|^2-|d_{\mu_i}(m)|^2$ takes the form,
\begin{equation}
\!\!|d_{\mu}(n)|^2\!-\!|d_{\mu}(m)|^2=2\epsilon{\rm Im}\Big[\la \mu|\hat{A}|\Psi_A\ra\la \Psi_A|\mu\ra\Big](n-m).\label{eq:dmu_general}
\end{equation}
Therefore, in the weak system–ancilla coupling limit, the time scale of global symmetry restoration is once again governed by the smallest separation between the number sectors present in the superposition. Consequently, even in the general case, initial states containing superpositions of distant number sectors restore symmetry faster than those involving nearby sectors.

\vspace{0.2cm}

\textit{Effect of local monitoring.--}
\begin{figure}
    \centering
    \includegraphics[width=0.52\linewidth,height=2in]{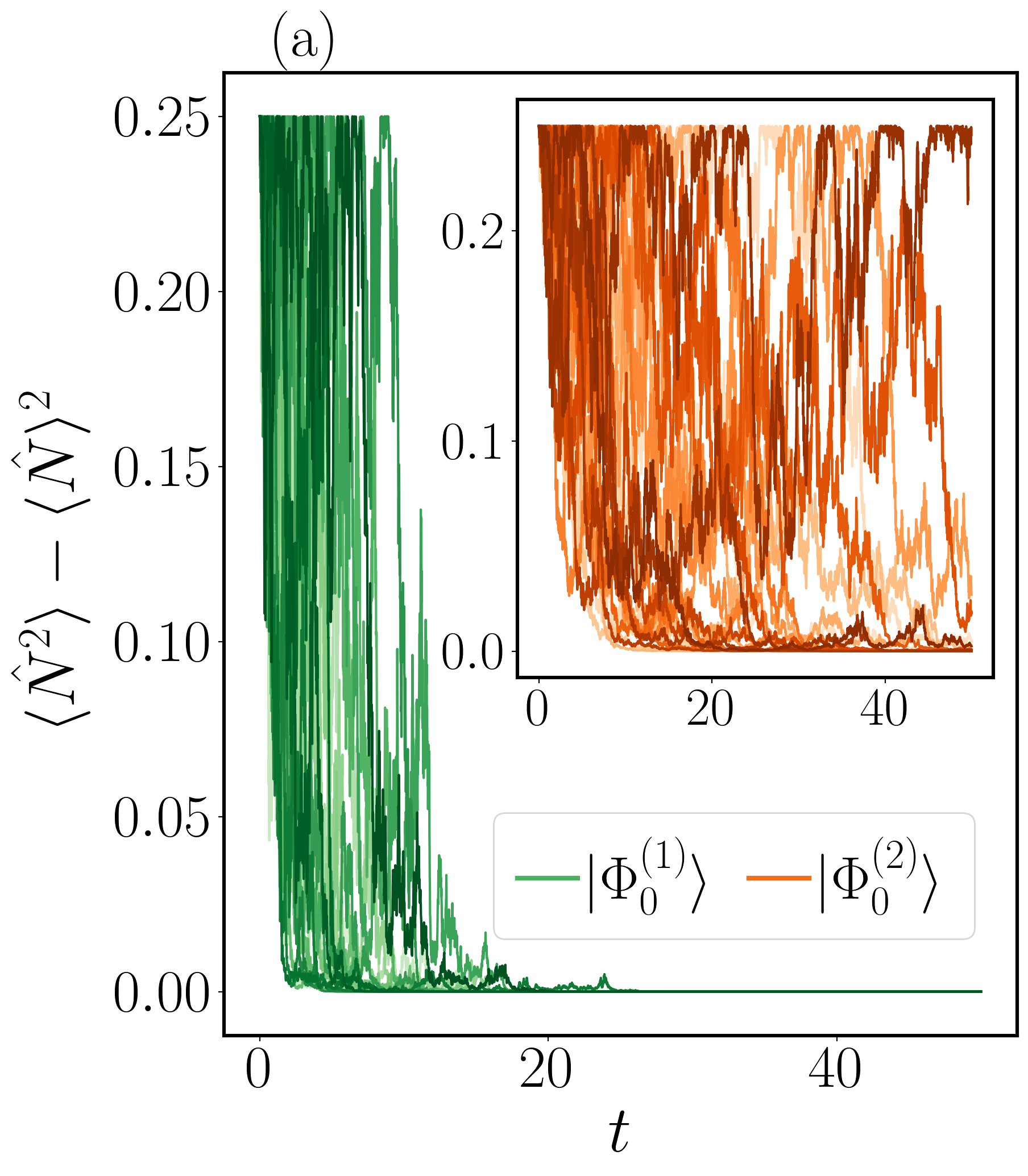}%
    \includegraphics[width=0.5\linewidth]{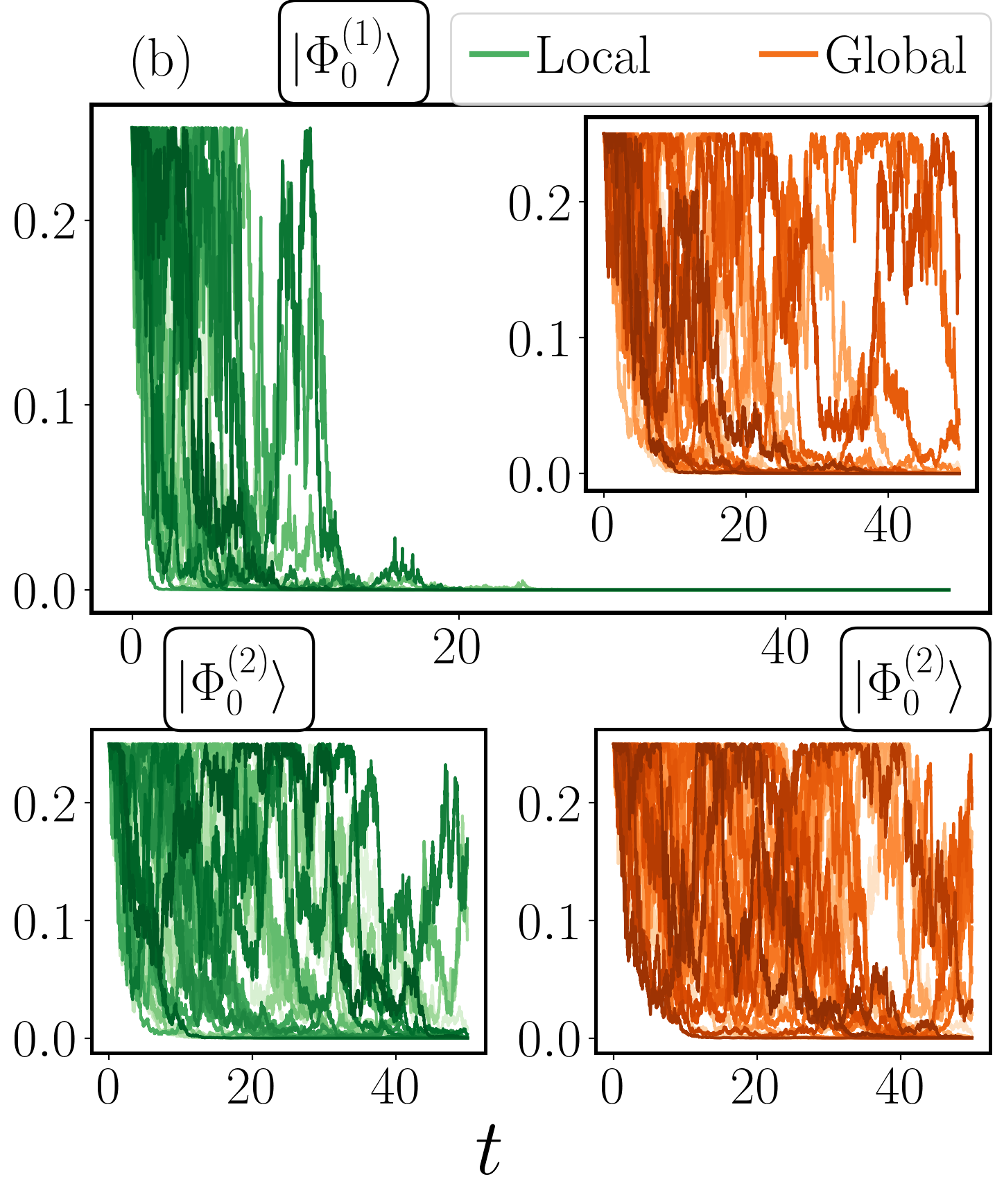}
    \caption{Plot of total number fluctuations under local monitoring of operator $\hat{n}_i$ under QSD protocol [Eq.~\eqref{eq:QSD_local_n}]. The parameters used here: $L=5$, $\gamma=0.1$, $dt=0.01$, unitary part through XX Hamiltonian with $J=0.001$. We have plotted $100$ different trajectories. (a) Number fluctuation is plotted for two different initial states $|\Phi_0^{(1)}\ra$ (main figure) and $|\Phi_0^{(2)}\ra$ (inset). $|\Phi_0^{(1)}\ra$ is restoring the symmetry faster than $|\Phi_0^{(2)}\ra$. (b) We have compared the symmetry restoration in local (green) and global (brown) monitoring for initial states $|\Phi_0^{(1)}\ra$ (top panel) and $|\Phi_0^{(2)}\ra$ (bottom panel). The top panel shows that, for the initial state $|\Phi_0^{(1)}\ra$, local monitoring (green) leads to a faster restoration of symmetry than global monitoring (brown). The bottom panel shows that, for the initial state $|\Phi_0^{(2)}\ra$, both local (green) and global (brown) monitoring restores symmetry in equivalent timescales.}
    \label{fig:local_monitoring}
\end{figure}
We now demonstrate the dynamical restoration of global $U(1)$ symmetry induced by continuous monitoring of the local operator $\hat{n}_i=\hat{S}_z^i+1/2$ throughout the lattice.
We focus on the quantum state diffusion (QSD) protocol. We will show that, in the presence of local monitoring, certain states that exhibit slow symmetry restoration under global monitoring instead undergo significantly faster symmetry restoration.
The SSE under QSD protocol takes the form,
\begin{align}
    d|\Psi_t\rangle\! =\!\Big[\!\! -i\hat{H}_S dt\,-\,&\frac{\gamma dt}{2}\sum_{i=1}^L(\hat{n}_i\!-\!\langle\hat{n}_i\rangle_t)^2\nonumber\\&+\sum_{i=1}^L(\hat{n}_i\!-\!\langle\hat{n}_i\rangle_t)\, d\xi^i_t\Big]|\Psi_t\ra, \label{eq:QSD_local_n}
\end{align}
where $\gamma$ is the strength of the monitoring at each site $i$ and $d\xi_t^i$ is the Wiener increment with the statistics $\overline{d\xi_t^i}=0$ and $\overline{d\xi_t^i \,d\xi_t^j}=\gamma dt\delta_{ij}$. Unlike global monitoring, under local monitoring the Hamiltonian plays a crucial role in symmetry restoration, as it does not commute with the local density operator $\hat{n}_i$. Here, once again, we take $H_S$ to be the Hamiltonian of XX spin chain. In the strong-measurement regime $J<\!\!<\gamma$, for any finite lattice size, symmetry restoration is guaranteed: since $\hat{n}_i$ commutes with the operator $\hat{N}$, and the Hamiltonian contribution is parametrically weaker than the monitoring-induced dynamics.
We again start from a $U(1)$ broken initial state $|\Psi_0\ra=\sum_n c^0_n|n\ra_0$ where $c_n^0$ denotes probability amplitude associated with $n$-th number sector, and $|n\ra_0$ is a state within that sector, which may in general be a superposition of different configurations. Under local monitoring, the states $|n\ra_0$ evolves to $|n\ra_t$. Therefore, the change in probability $|c_n^t|^2=|\la\Psi_t|n\ra_t|^2$ under the local monitoring follows the equation
\begin{equation}
    d|c_n^t|^2=|c_n^t|^2\sum_{m\neq n}\sum_{i=1}^{L}d\xi_t^i\Big[{_t}\la n|\hat{n}_i|n\ra_t-{_t}\la m|\hat{n}_i|m\ra_t\Big]. \label{eq:prob_local_monitoring}
\end{equation}
Interestingly, Eq.~\eqref{eq:prob_local_monitoring} shows that the symmetry-restoration dynamics is governed by differences in the local density profiles between distinct number sectors, ${_t}\la n|\hat{n}_i|n\ra_t-{_t}\la m|\hat{n}_i|m\ra_t$. More precisely, since the Wiener increments $d\xi_t^i$,
are independent Gaussian variables, the stochastic term
$\sum
_{i} d\xi_t^i\big[{_t}\la n|\hat{n}_i|n\ra_t-{_t}\la m|\hat{n}_i|m\ra_t\big]$is itself a Gaussian noise with zero mean and variance $\gamma dt\sum_i\big[{_t}\la n|\hat{n}_i|n\ra_t-{_t}\la m|\hat{n}_i|m\ra_t\big]^2$. This implies that, for an initial state containing superpositions of different number sectors, the time scale of symmetry restoration is controlled by the smallest separation between the corresponding local density profiles. 
In other words, an initial state with smaller overlap between the density profiles across different number sectors exhibits faster symmetry restoration than states with larger overlap.

To further illustrate this point, we consider the same class of initial states used previously in the context of global monitoring. 
The distant number sectors typically maintain large differences in density profiles and hence under local monitoring, those states again restore the symmetry quickly. However, the states with superposition of nearby sectors that always show slower symmetry restoration under global monitoring, can now become faster under local monitoring for certain initial configurations. We consider two such different initial configurations 
$|\Phi_0^{(1)}\ra=\big[|\!\!\uparrow\downarrow\uparrow\downarrow\uparrow \ra+|\!\!\downarrow\uparrow\downarrow\uparrow\downarrow\ra\big)/\sqrt{2}$ and $|\Phi_0^{(2)}\ra=\big[|\!\!\uparrow\uparrow\uparrow\downarrow\downarrow \ra+|\!\!\uparrow\uparrow\uparrow\uparrow\downarrow\ra\big)/\sqrt{2}$. Note that in the state $|\Phi_0^{(1)}\ra$, the density profiles of the two number sectors do not have any overlap whereas in $|\Phi_0^{(2)}\ra$, the density profiles have high overlap. Consequently, the initial state $|\Phi_0^{(1)}\ra$ restores the symmetry much faster than $|\Phi_0^{(2)}\ra$. In Fig.~\ref{fig:local_monitoring}, we have illustrated our numerical results of local monitoring: Fig.~\ref{fig:local_monitoring}(a) represents the plot of fluctuations in $\hat{N}$ with time for the initial states $|\Phi_0^{(1)}\ra$ and $|\Phi_0^{(2)}\ra$. As expected from the previous discussion, the trajectory originating from $|\Phi_0^{(1)}\ra$ restores the symmetry faster than trajectory generated from $|\Phi_0^{(2)}\ra$. In Fig.~\ref{fig:local_monitoring}(b), we compare the symmetry restoration dynamics under local and global monitoring. We observe that for the state $|\Phi_0^{(1)}\ra$ [top panel of Fig.~\ref{fig:local_monitoring}(b)], local monitoring always leads to faster relaxation than global monitoring. In contrast, for the state $|\Phi_0^{(2)}\ra$, symmetry restoration timescales are equivalent under local and global monitoring.



\vspace{0.2cm}
\textit{Summary.--}
Accelerated relaxation in quantum dynamics is of broad interest in a variety of contexts, including processes involving repeated resetting~\cite{Bao2025,solanki2025}, accelerated dissipative state preparation~\cite{Clerk2025}, and fast initialization of quantum devices~\cite{Tuorila2017,Liu_2025}. In this work, we achieve accelerated symmetry restoration  in quantum trajectories by utilizing measurement backaction. We first consider continuous monitoring of global number operator $\hat{N}$ combined with $U(1)$ conserving unitary dynamics. Here, we observe that the time-scale of symmetry restoration depends on the minimum separation between the number sectors present in the initial superposition. The initial states involving superposition of distant number sectors restores the global $U(1)$ symmetry faster than states involving nearby number sectors which universally holds for different measurement protocol. However, under the continuous monitoring of the local operator $\hat{n}_i$, we observe that time-scale of symmetry restoration depends on overlap between the density profiles of different number sectors present in the initial state. The states involving less overlap between density profiles across different number sectors relax much faster than any other states, irrespective of separation between the charge sectors. Hence, few states that show slower symmetry restoration under global monitoring, can be accelerated through local monitoring.
Overall, our study unveils that by appropriately tailoring global and local continuous measurement, one can significantly speed up relaxation dynamics in quantum trajectories. 

In future, it will be interesting to generalize the observed symmetry restoration for systems with multiple conserved charges that are not commutating. The effect of Hamiltonian in the local measurement would also be another interesting future outlook.

\vspace{0.2cm}
\textit{Acknowledgements.--}  
KG acknowledges Sandipan Manna for useful discussions.
BKA acknowledges the CRG grant No. CRG/2023/003377 from ANRF, Government of India. KG would like to acknowledge the Prime Minister's Research Fellowship (ID- 0703043), Government of India for funding. KG and BKA acknowledge the National Supercomputing Mission (NSM) for providing computing resources of ‘PARAM Brahma’ at IISER Pune, which is implemented by C-DAC and supported by the Ministry of Electronics and Information Technology (MeitY) and DST, Government of India.

\bibliography{references.bib}

@article{Adolfo_2019,
  title = {Spontaneous Symmetry Breaking Induced by Quantum Monitoring},
  author = {Garc\'{\i}a-Pintos, Luis Pedro and Tielas, Diego and del Campo, Adolfo},
  journal = {Phys. Rev. Lett.},
  volume = {123},
  issue = {9},
  pages = {090403},
  numpages = {6},
  year = {2019},
  month = {Aug},
  publisher = {American Physical Society},
  doi = {10.1103/PhysRevLett.123.090403},
  url = {https://link.aps.org/doi/10.1103/PhysRevLett.123.090403}
}

@article{Sarang_2022,
  title = {Entanglement and Charge-Sharpening Transitions in U(1) Symmetric Monitored Quantum Circuits},
  author = {Agrawal, Utkarsh and Zabalo, Aidan and Chen, Kun and Wilson, Justin H. and Potter, Andrew C. and Pixley, J. H. and Gopalakrishnan, Sarang and Vasseur, Romain},
  journal = {Phys. Rev. X},
  volume = {12},
  issue = {4},
  pages = {041002},
  numpages = {29},
  year = {2022},
  month = {Oct},
  publisher = {American Physical Society},
  doi = {10.1103/PhysRevX.12.041002},
  url = {https://link.aps.org/doi/10.1103/PhysRevX.12.041002}
}

@article{Nahum_2019,
  title = {Measurement-Induced Phase Transitions in the Dynamics of Entanglement},
  author = {Skinner, Brian and Ruhman, Jonathan and Nahum, Adam},
  journal = {Phys. Rev. X},
  volume = {9},
  issue = {3},
  pages = {031009},
  numpages = {21},
  year = {2019},
  month = {Jul},
  publisher = {American Physical Society},
  doi = {10.1103/PhysRevX.9.031009},
  url = {https://link.aps.org/doi/10.1103/PhysRevX.9.031009}
}

@article{VG1976,
  title={Completely positive dynamical semigroups of N-level systems},
  author={Gorini, Vittorio and Kossakowski, Andrzej and Sudarshan, Ennackal Chandy George},
  journal={Journal of Mathematical Physics},
  volume={17},
  number={5},
  pages={821--825},
  year={1976},
  publisher={American Institute of Physics},
  url={https://doi.org/10.1063/1.522979}
}

@article{L1976,
  title={On the generators of quantum dynamical semigroups},
  author={Lindblad, Goran},
  journal={Communications in mathematical physics},
  volume={48},
  number={2},
  pages={119--130},
  year={1976},
  publisher={Springer},
  url = {https://projecteuclid.org/journals/communications-in-mathematical-physics/volume-48/issue-2/On-the-generators-of-quantum-dynamical-semigroups/cmp/1103899849.full}
}

@book{BPOQS,
    author = {Breuer, Heinz-Peter and Petruccione, Francesco},
    title = {The Theory of Open Quantum Systems},
    publisher = {Oxford University Press},
    year = {2007},
    month = {01},
    isbn = {9780199213900},
    doi = {10.1093/acprof:oso/9780199213900.001.0001},
    url = {https://doi.org/10.1093/acprof:oso/9780199213900.001.0001},
}

@book{carmichael2009,
  title={An Open Systems Approach to Quantum Optics: Lectures Presented at the Universit{\'e} Libre de Bruxelles, October 28 to November 4, 1991},
  author={Carmichael, H.},
  isbn={9783540476207},
  lccn={93019394},
  series={Lecture Notes in Physics Monographs},
  url={https://books.google.co.in/books?id=uor_CAAAQBAJ},
  year={2009},
  publisher={Springer Berlin Heidelberg}
}

@book{carmichael1998,
  title={Statistical Methods in Quantum Optics 1: Master Equations and Fokker-Planck Equations},
  author={Carmichael, H.},
  isbn={9783540548829},
  lccn={98040873},
  series={Physics and astronomy online library},
  url={https://books.google.co.in/books?id=ocgRgM-yJacC},
  year={1998},
  publisher={Springer}
}

@book{wiseman2010,
  title={Quantum Measurement and Control},
  author={Wiseman, H.M. and Milburn, G.J.},
  isbn={9780521804424},
  lccn={2009034266},
  url={https://books.google.co.in/books?id=ZNjvHaH8qA4C},
  year={2010},
  publisher={Cambridge University Press}
}

@book{jacobs2014,
  title={Quantum Measurement Theory and its Applications},
  author={Jacobs, K.},
  isbn={9781107025486},
  lccn={2014011297},
  url={https://books.google.co.in/books?id=gzTRngEACAAJ},
  year={2014},
  publisher={Cambridge University Press}
}

@article{Russomano_2022,
  title = {Entanglement transitions in the quantum Ising chain: A comparison between different unravelings of the same Lindbladian},
  author = {Piccitto, Giulia and Russomanno, Angelo and Rossini, Davide},
  journal = {Phys. Rev. B},
  volume = {105},
  issue = {6},
  pages = {064305},
  numpages = {14},
  year = {2022},
  month = {Feb},
  publisher = {American Physical Society},
  doi = {10.1103/PhysRevB.105.064305},
  url = {https://link.aps.org/doi/10.1103/PhysRevB.105.064305}
}

@article{Kater2013,
	author = {Murch, K. W. and Weber, S. J. and Macklin, C. and Siddiqi, I.},
	journal = {Nature},
	number = {7470},
	title = {Observing single quantum trajectories of a superconducting quantum bit},
	url = {https://doi.org/10.1038/nature12539},
	volume = {502},
	year = {2013}}

@article{Rainer1986,
  title = {Observation of Quantum Jumps},
  author = {Sauter, Th. and Neuhauser, W. and Blatt, R. and Toschek, P. E.},
  journal = {Phys. Rev. Lett.},
  volume = {57},
  issue = {14},
  pages = {1696--1698},
  year = {1986},
  month = {Oct},
  publisher = {American Physical Society},
  doi = {10.1103/PhysRevLett.57.1696},
  url = {https://link.aps.org/doi/10.1103/PhysRevLett.57.1696}
}

@article{Vijay2011,
  title = {Observation of Quantum Jumps in a Superconducting Artificial Atom},
  author = {Vijay, R. and Slichter, D. H. and Siddiqi, I.},
  journal = {Phys. Rev. Lett.},
  volume = {106},
  issue = {11},
  pages = {110502},
  numpages = {4},
  year = {2011},
  month = {Mar},
  publisher = {American Physical Society},
  doi = {10.1103/PhysRevLett.106.110502},
  url = {https://link.aps.org/doi/10.1103/PhysRevLett.106.110502}
}

@article{Sebastian2021,
  title = {Entanglement Transition in a Monitored Free-Fermion Chain: From Extended Criticality to Area Law},
  author = {Alberton, O. and Buchhold, M. and Diehl, S.},
  journal = {Phys. Rev. Lett.},
  volume = {126},
  issue = {17},
  pages = {170602},
  numpages = {6},
  year = {2021},
  month = {Apr},
  publisher = {American Physical Society},
  doi = {10.1103/PhysRevLett.126.170602},
  url = {https://link.aps.org/doi/10.1103/PhysRevLett.126.170602}
}

@article{Saito2022,
  title = {Fate of Measurement-Induced Phase Transition in Long-Range Interactions},
  author = {Minato, Takaaki and Sugimoto, Koudai and Kuwahara, Tomotaka and Saito, Keiji},
  journal = {Phys. Rev. Lett.},
  volume = {128},
  issue = {1},
  pages = {010603},
  numpages = {7},
  year = {2022},
  month = {Jan},
  publisher = {American Physical Society},
  doi = {10.1103/PhysRevLett.128.010603},
  url = {https://link.aps.org/doi/10.1103/PhysRevLett.128.010603}
}

@article{Muller2022,
  title = {Measurement-Induced Dark State Phase Transitions in Long-Ranged Fermion Systems},
  author = {M\"uller, T. and Diehl, S. and Buchhold, M.},
  journal = {Phys. Rev. Lett.},
  volume = {128},
  issue = {1},
  pages = {010605},
  numpages = {6},
  year = {2022},
  month = {Jan},
  publisher = {American Physical Society},
  doi = {10.1103/PhysRevLett.128.010605},
  url = {https://link.aps.org/doi/10.1103/PhysRevLett.128.010605}
}

@article{Khemani-Huse2021,
  title = {Entanglement Phase Transitions in Measurement-Only Dynamics},
  author = {Ippoliti, Matteo and Gullans, Michael J. and Gopalakrishnan, Sarang and Huse, David A. and Khemani, Vedika},
  journal = {Phys. Rev. X},
  volume = {11},
  issue = {1},
  pages = {011030},
  numpages = {23},
  year = {2021},
  month = {Feb},
  publisher = {American Physical Society},
  doi = {10.1103/PhysRevX.11.011030},
  url = {https://link.aps.org/doi/10.1103/PhysRevX.11.011030}
}

@article{Romito2019,
  title = {Entanglement transition from variable-strength weak measurements},
  author = {Szyniszewski, M. and Romito, A. and Schomerus, H.},
  journal = {Phys. Rev. B},
  volume = {100},
  issue = {6},
  pages = {064204},
  numpages = {8},
  year = {2019},
  month = {Aug},
  publisher = {American Physical Society},
  doi = {10.1103/PhysRevB.100.064204},
  url = {https://link.aps.org/doi/10.1103/PhysRevB.100.064204}
}

@article{Fazio2025,
  title = {Measurement-induced phase transitions in monitored infinite-range interacting systems},
  author = {Delmonte, Anna and Li, Zejian and Passarelli, Gianluca and Song, Eric Yilun and Barberena, Diego and Rey, Ana Maria and Fazio, Rosario},
  journal = {Phys. Rev. Res.},
  volume = {7},
  issue = {2},
  pages = {023082},
  numpages = {27},
  year = {2025},
  month = {Apr},
  publisher = {American Physical Society},
  doi = {10.1103/PhysRevResearch.7.023082},
  url = {https://link.aps.org/doi/10.1103/PhysRevResearch.7.023082}
}

@article{Andrew2022,
  title = {Transitions in the Learnability of Global Charges from Local Measurements},
  author = {Barratt, Fergus and Agrawal, Utkarsh and Potter, Andrew C. and Gopalakrishnan, Sarang and Vasseur, Romain},
  journal = {Phys. Rev. Lett.},
  volume = {129},
  issue = {20},
  pages = {200602},
  numpages = {7},
  year = {2022},
  month = {Nov},
  publisher = {American Physical Society},
  doi = {10.1103/PhysRevLett.129.200602},
  url = {https://link.aps.org/doi/10.1103/PhysRevLett.129.200602}
}

@article{Khemani2024,
  title = {Learnability Transitions in Monitored Quantum Dynamics via Eavesdropper's Classical Shadows},
  author = {Ippoliti, Matteo and Khemani, Vedika},
  journal = {PRX Quantum},
  volume = {5},
  issue = {2},
  pages = {020304},
  numpages = {24},
  year = {2024},
  month = {Apr},
  publisher = {American Physical Society},
  doi = {10.1103/PRXQuantum.5.020304},
  url = {https://link.aps.org/doi/10.1103/PRXQuantum.5.020304}
}

@article{Phillip2025,
  title = {Symmetry-protection Zeno phase transition in monitored lattice gauge theories},
  author = {Wauters, Matteo M. and Ballini, Edoardo and Biella, Alberto and Hauke, Philipp},
  journal = {Phys. Rev. B},
  volume = {111},
  issue = {9},
  pages = {094315},
  numpages = {18},
  year = {2025},
  month = {Mar},
  publisher = {American Physical Society},
  doi = {10.1103/PhysRevB.111.094315},
  url = {https://link.aps.org/doi/10.1103/PhysRevB.111.094315}
}

@article{MBPJan1998,
  title = {The quantum-jump approach to dissipative dynamics in quantum optics},
  author = {Plenio, M. B. and Knight, P. L.},
  journal = {Rev. Mod. Phys.},
  volume = {70},
  issue = {1},
  pages = {101--144},
  numpages = {0},
  year = {1998},
  month = {Jan},
  publisher = {American Physical Society},
  doi = {10.1103/RevModPhys.70.101},
  url = {https://link.aps.org/doi/10.1103/RevModPhys.70.101}
}

@article{IROct2015,
doi = {10.1088/0034-4885/78/11/114001},
url = {https://dx.doi.org/10.1088/0034-4885/78/11/114001},
year = {2015},
month = {oct},
publisher = {IOP Publishing},
volume = {78},
number = {11},
pages = {114001},
author = {Rotter, I and Bird, J P},
title = {A review of progress in the physics of open quantum systems: theory and experiment},
journal = {Reports on Progress in Physics}
}

@article{AMJFeb1982,
  title = {Nondiffusive Quantum Transport in a Dynamically Disordered Medium},
  author = {Jayannavar, A. M. and Kumar, N.},
  journal = {Phys. Rev. Lett.},
  volume = {48},
  issue = {8},
  pages = {553--556},
  numpages = {0},
  year = {1982},
  month = {Feb},
  publisher = {American Physical Society},
  doi = {10.1103/PhysRevLett.48.553},
  url = {https://link.aps.org/doi/10.1103/PhysRevLett.48.553}
}

@article{HPBApr2016,
  title = {Colloquium: Non-Markovian dynamics in open quantum systems},
  author = {Breuer, Heinz-Peter and Laine, Elsi-Mari and Piilo, Jyrki and Vacchini, Bassano},
  journal = {Rev. Mod. Phys.},
  volume = {88},
  issue = {2},
  pages = {021002},
  numpages = {24},
  year = {2016},
  month = {Apr},
  publisher = {American Physical Society},
  doi = {10.1103/RevModPhys.88.021002},
  url = {https://link.aps.org/doi/10.1103/RevModPhys.88.021002}
}

@article{Aurelia2025,
  title = {Quantum Dynamics with Stochastic Non-Hermitian Hamiltonians},
  author = {Martinez-Azcona, Pablo and Kundu, Aritra and Saxena, Avadh and del Campo, Adolfo and Chenu, Aur\'elia},
  journal = {Phys. Rev. Lett.},
  volume = {135},
  issue = {1},
  pages = {010402},
  numpages = {6},
  year = {2025},
  month = {Jul},
  publisher = {American Physical Society},
  doi = {10.1103/5ksl-tjjm},
  url = {https://link.aps.org/doi/10.1103/5ksl-tjjm}
}

@article{PN2025,
    title = {Quantum dynamics in Krylov space: Methods and applications},
    journal = {Physics Reports},
    volume = {1125-1128},
    pages = {1-82},
    year = {2025},
    doi = {https://doi.org/10.1016/j.physrep.2025.05.001},
    url = {https://www.sciencedirect.com/science/article/pii/S0370157325001462},
    author = {Pratik Nandy and Apollonas S. Matsoukas-Roubeas and Pablo Martínez-Azcona and Anatoly Dymarsky and Adolfo {del Campo}},
}

@article{Shiwu1998,
  title = {Lindblad approach to quantum dynamics of open systems},
  author = {Gao, Shiwu},
  journal = {Phys. Rev. B},
  volume = {57},
  issue = {8},
  pages = {4509--4517},
  numpages = {0},
  year = {1998},
  month = {Feb},
  publisher = {American Physical Society},
  doi = {10.1103/PhysRevB.57.4509},
  url = {https://link.aps.org/doi/10.1103/PhysRevB.57.4509}
}

@article{Kapral_2015,
doi = {10.1088/0953-8984/27/7/073201},
url = {https://doi.org/10.1088/0953-8984/27/7/073201},
year = {2015},
month = {jan},
publisher = {IOP Publishing},
volume = {27},
number = {7},
pages = {073201},
author = {Kapral, Raymond},
title = {Quantum dynamics in open quantum-classical systems},
journal = {Journal of Physics: Condensed Matter}}

@article{Mariia2025,
	author = {Ivanchenko, Mariia and Walters, Peter L. and Wang, Fei},
	journal = {Journal of Chemical Theory and Computation},
	number = {12},
	title = {Investigating Non-Markovian Effects on Quantum Dynamics in Open Quantum Systems},
	url = {https://doi.org/10.1021/acs.jctc.4c01632},
	volume = {21},
	year = {2025}}

@article{dutta2025,
      title={An introduction to Markovian open quantum systems}, 
      author={Shovan Dutta},
      year={2025},
      journal = {arXiv:2510.26530},
      url={https://arxiv.org/abs/2510.26530}, 
}

@Article{Sciro2025,
	title={{Many-body open quantum systems}},
	author={Rosario Fazio and Jonathan Keeling and Leonardo Mazza and Marco Schirò},
	journal={SciPost Phys. Lect. Notes},
	pages={99},
	year={2025},
	publisher={SciPost},
	doi={10.21468/SciPostPhysLectNotes.99},
	url={https://scipost.org/10.21468/SciPostPhysLectNotes.99},
}

@article{Daniel2017,
  title = {Dynamics of non-Markovian open quantum systems},
  author = {de Vega, In\'es and Alonso, Daniel},
  journal = {Rev. Mod. Phys.},
  volume = {89},
  issue = {1},
  pages = {015001},
  numpages = {58},
  year = {2017},
  month = {Jan},
  publisher = {American Physical Society},
  doi = {10.1103/RevModPhys.89.015001},
  url = {https://link.aps.org/doi/10.1103/RevModPhys.89.015001}
}

@article{Plenio2018,
  title = {Nonperturbative Treatment of non-Markovian Dynamics of Open Quantum Systems},
  author = {Tamascelli, D. and Smirne, A. and Huelga, S. F. and Plenio, M. B.},
  journal = {Phys. Rev. Lett.},
  volume = {120},
  issue = {3},
  pages = {030402},
  numpages = {6},
  year = {2018},
  month = {Jan},
  publisher = {American Physical Society},
  doi = {10.1103/PhysRevLett.120.030402},
  url = {https://link.aps.org/doi/10.1103/PhysRevLett.120.030402}
}

@article{Franco2011,
  title = {Quantum algorithm for simulating the dynamics of an open quantum system},
  author = {Wang, Hefeng and Ashhab, S. and Nori, Franco},
  journal = {Phys. Rev. A},
  volume = {83},
  issue = {6},
  pages = {062317},
  numpages = {11},
  year = {2011},
  month = {Jun},
  publisher = {American Physical Society},
  doi = {10.1103/PhysRevA.83.062317},
  url = {https://link.aps.org/doi/10.1103/PhysRevA.83.062317}
}

@article{Yan2002,
    author = {Xu, Ruixue and Yan, YiJing},
    title = { Theory of open quantum systems},
    journal = {The Journal of Chemical Physics},
    volume = {116},
    number = {21},
    year = {2002},
    url = {https://doi.org/10.1063/1.1474579}
}

@article{Dries2020,
  title = {Non-Gaussian correlations imprinted by local dephasing in fermionic wires},
  author = {Dolgirev, Pavel E. and Marino, Jamir and Sels, Dries and Demler, Eugene},
  journal = {Phys. Rev. B},
  volume = {102},
  issue = {10},
  pages = {100301},
  numpages = {6},
  year = {2020},
  month = {Sep},
  publisher = {American Physical Society},
  doi = {10.1103/PhysRevB.102.100301},
  url = {https://link.aps.org/doi/10.1103/PhysRevB.102.100301}
}

@article{Adolfo2017,
  title = {Quantum Simulation of Generic Many-Body Open System Dynamics Using Classical Noise},
  author = {Chenu, A. and Beau, M. and Cao, J. and del Campo, A.},
  journal = {Phys. Rev. Lett.},
  volume = {118},
  issue = {14},
  pages = {140403},
  numpages = {6},
  year = {2017},
  month = {Apr},
  publisher = {American Physical Society},
  doi = {10.1103/PhysRevLett.118.140403},
  url = {https://link.aps.org/doi/10.1103/PhysRevLett.118.140403}
}

@article{Zhong2019,
  title = {Non-Hermitian Skin Effect and Chiral Damping in Open Quantum Systems},
  author = {Song, Fei and Yao, Shunyu and Wang, Zhong},
  journal = {Phys. Rev. Lett.},
  volume = {123},
  issue = {17},
  pages = {170401},
  numpages = {8},
  year = {2019},
  month = {Oct},
  publisher = {American Physical Society},
  doi = {10.1103/PhysRevLett.123.170401},
  url = {https://link.aps.org/doi/10.1103/PhysRevLett.123.170401}
}

@article{Prosen2023,
  title = {Anomalous Diffusion in the Long-Range Haken-Strobl-Reineker Model},
  author = {Catalano, A. G. and Mattiotti, F. and Dubail, J. and Hagenm\"uller, D. and Prosen, T. and Franchini, F. and Pupillo, G.},
  journal = {Phys. Rev. Lett.},
  volume = {131},
  issue = {5},
  pages = {053401},
  numpages = {6},
  year = {2023},
  month = {Aug},
  publisher = {American Physical Society},
  doi = {10.1103/PhysRevLett.131.053401},
  url = {https://link.aps.org/doi/10.1103/PhysRevLett.131.053401}
}

@article{SGJul2017,
  title = {Noise-Induced Subdiffusion in Strongly Localized Quantum Systems},
  author = {Gopalakrishnan, Sarang and Islam, K. Ranjibul and Knap, Michael},
  journal = {Phys. Rev. Lett.},
  volume = {119},
  issue = {4},
  pages = {046601},
  numpages = {6},
  year = {2017},
  month = {Jul},
  publisher = {American Physical Society},
  doi = {10.1103/PhysRevLett.119.046601},
  url = {https://link.aps.org/doi/10.1103/PhysRevLett.119.046601}
}

@article{Archak2018,
  title = {Anomalous transport in the Aubry-Andr\'e-Harper model in isolated and open systems},
  author = {Purkayastha, Archak and Sanyal, Sambuddha and Dhar, Abhishek and Kulkarni, Manas},
  journal = {Phys. Rev. B},
  volume = {97},
  issue = {17},
  pages = {174206},
  numpages = {11},
  year = {2018},
  month = {May},
  publisher = {American Physical Society},
  doi = {10.1103/PhysRevB.97.174206},
  url = {https://link.aps.org/doi/10.1103/PhysRevB.97.174206}
}

@article{LBMar2015,
  title = {Absence of Diffusion in an Interacting System of Spinless Fermions on a One-Dimensional Disordered Lattice},
  author = {Bar Lev, Yevgeny and Cohen, Guy and Reichman, David R.},
  journal = {Phys. Rev. Lett.},
  volume = {114},
  issue = {10},
  pages = {100601},
  numpages = {5},
  year = {2015},
  month = {Mar},
  publisher = {American Physical Society},
  doi = {10.1103/PhysRevLett.114.100601},
  url = {https://link.aps.org/doi/10.1103/PhysRevLett.114.100601}
}

@article{JSBJan2018,
  title = {Light-Cone and Diffusive Propagation of Correlations in a Many-Body Dissipative System},
  author = {Bernier, Jean-S\'ebastien and Tan, Ryan and Bonnes, Lars and Guo, Chu and Poletti, Dario and Kollath, Corinna},
  journal = {Phys. Rev. Lett.},
  volume = {120},
  issue = {2},
  pages = {020401},
  numpages = {5},
  year = {2018},
  month = {Jan},
  publisher = {American Physical Society},
  doi = {10.1103/PhysRevLett.120.020401},
  url = {https://link.aps.org/doi/10.1103/PhysRevLett.120.020401}
}

@Article{DSB2024,
	title={{Noise-induced transport in the Aubry-André-Harper model}},
	author={Devendra Singh Bhakuni and  Talía L. M. Lezama and Yevgeny Bar Lev},
	journal={SciPost Phys. Core},
	volume={7},
	pages={023},
	year={2024},
	publisher={SciPost},
	doi={10.21468/SciPostPhysCore.7.2.023},
	url={https://scipost.org/10.21468/SciPostPhysCore.7.2.023},
}

@Article{YPW2024,
	title={{Superdiffusive transport in quasi-particle dephasing models}},
	author={Yu-Peng Wang and Chen Fang and Jie Ren},
	journal={SciPost Phys.},
	volume={17},
	pages={150},
	year={2024},
	publisher={SciPost},
	doi={10.21468/SciPostPhys.17.6.150},
	url={https://scipost.org/10.21468/SciPostPhys.17.6.150},
}

@article{NGisin_1992,
doi = {10.1088/0305-4470/25/21/023},
url = {https://doi.org/10.1088/0305-4470/25/21/023},
year = {1992},
month = {nov},
publisher = {},
volume = {25},
number = {21},
pages = {5677},
author = {N Gisin and I C Percival},
title = {The quantum-state diffusion model applied to open systems},
journal = {Journal of Physics A: Mathematical and General}}

@article{Molmer2008,
  title = {Wave-function approach to dissipative processes in quantum optics},
  author = {Dalibard, Jean and Castin, Yvan and M\o{}lmer, Klaus},
  journal = {Phys. Rev. Lett.},
  volume = {68},
  issue = {5},
  pages = {580--583},
  numpages = {0},
  year = {1992},
  month = {Feb},
  publisher = {American Physical Society},
  doi = {10.1103/PhysRevLett.68.580},
  url = {https://link.aps.org/doi/10.1103/PhysRevLett.68.580}
}

@article{GISIN1992315,
title = {Wave-function approach to dissipative processes: are there quantum jumps?},
journal = {Physics Letters A},
volume = {167},
number = {4},
pages = {315-318},
year = {1992},
issn = {0375-9601},
doi = {https://doi.org/10.1016/0375-9601(92)90264-M},
url = {https://www.sciencedirect.com/science/article/pii/037596019290264M},
author = {Nicolas Gisin and Ian C. Percival},
}

@article{Blatt2019,
  title = {Environment-Assisted Quantum Transport in a 10-qubit Network},
  author = {Maier, Christine and Brydges, Tiff and Jurcevic, Petar and Trautmann, Nils and Hempel, Cornelius and Lanyon, Ben P. and Hauke, Philipp and Blatt, Rainer and Roos, Christian F.},
  journal = {Phys. Rev. Lett.},
  volume = {122},
  issue = {5},
  pages = {050501},
  numpages = {6},
  year = {2019},
  month = {Feb},
  publisher = {American Physical Society},
  doi = {10.1103/PhysRevLett.122.050501},
  url = {https://link.aps.org/doi/10.1103/PhysRevLett.122.050501}
}

@article{WU2021213,
title = {Quantum computing and simulation with trapped ions: On the path to the future},
journal = {Fundamental Research},
volume = {1},
number = {2},
pages = {213-216},
year = {2021},
issn = {2667-3258},
doi = {https://doi.org/10.1016/j.fmre.2020.12.004},
url = {https://www.sciencedirect.com/science/article/pii/S2667325820300121},
author = {Wei Wu and Ting Zhang and Ping-Xing Chen}
}

@Article{Schäfer2020,
author={Sch{\"a}fer, Florian
and Fukuhara, Takeshi
and Sugawa, Seiji
and Takasu, Yosuke
and Takahashi, Yoshiro},
title={Tools for quantum simulation with ultracold atoms in optical lattices},
journal={Nature Reviews Physics},
year={2020},
month={Aug},
day={01},
volume={2},
number={8},
pages={411-425},
issn={2522-5820},
doi={10.1038/s42254-020-0195-3},
url={https://doi.org/10.1038/s42254-020-0195-3}
}

@Article{Bloch2012,
author={Bloch, Immanuel
and Dalibard, Jean
and Nascimb{\`e}ne, Sylvain},
title={Quantum simulations with ultracold quantum gases},
journal={Nature Physics},
year={2012},
month={Apr},
day={01},
volume={8},
number={4},
pages={267-276},
issn={1745-2481},
doi={10.1038/nphys2259},
url={https://doi.org/10.1038/nphys2259}
}

@Article{Blatt2012,
author={Blatt, R.
and Roos, C. F.},
title={Quantum simulations with trapped ions},
journal={Nature Physics},
year={2012},
month={Apr},
day={01},
volume={8},
number={4},
pages={277-284},
issn={1745-2481},
doi={10.1038/nphys2252},
url={https://doi.org/10.1038/nphys2252}
}

@article{Jacobs01092006,
author = {Kurt Jacobs and Daniel A. Steck},
title = {A straightforward introduction to continuous quantum measurement},
journal = {Contemporary Physics},
volume = {47},
number = {5},
pages = {279--303},
year = {2006},
publisher = {Taylor \& Francis},
doi = {10.1080/00107510601101934},
URL = { https://doi.org/10.1080/00107510601101934
}}

@article{Huse2020,
  title = {Dynamical Purification Phase Transition Induced by Quantum Measurements},
  author = {Gullans, Michael J. and Huse, David A.},
  journal = {Phys. Rev. X},
  volume = {10},
  issue = {4},
  pages = {041020},
  numpages = {28},
  year = {2020},
  month = {Oct},
  publisher = {American Physical Society},
  doi = {10.1103/PhysRevX.10.041020},
  url = {https://link.aps.org/doi/10.1103/PhysRevX.10.041020}
}

@article{Landi2024,
  title = {Current Fluctuations in Open Quantum Systems: Bridging the Gap Between Quantum Continuous Measurements and Full Counting Statistics},
  author = {Landi, Gabriel T. and Kewming, Michael J. and Mitchison, Mark T. and Potts, Patrick P.},
  journal = {PRX Quantum},
  volume = {5},
  issue = {2},
  pages = {020201},
  numpages = {86},
  year = {2024},
  month = {Apr},
  publisher = {American Physical Society},
  doi = {10.1103/PRXQuantum.5.020201},
  url = {https://link.aps.org/doi/10.1103/PRXQuantum.5.020201}
}

@book{mensky2017,
  title={Continuous Quantum Measurements and Path Integrals},
  author={Mensky, M.B.},
  isbn={9781351458023},
  url={https://books.google.co.in/books?id=dLs6DwAAQBAJ},
  year={2017},
  publisher={CRC Press}
}

@article{anzai2025,
      title={Disordered purification phase transition in hybrid random circuits}, 
      author={Kengo Anzai and Hiroaki Matsueda and Yoshihito Kuno},
      journal={arXiv:2507.12886},
      year={2025},
      url={https://arxiv.org/abs/2507.12886}, 
}

@article{Znidaric2015,
  title = {Relaxation times of dissipative many-body quantum systems},
  author = {\ifmmode \check{Z}\else \v{Z}\fi{}nidari\ifmmode \check{c}\else \v{c}\fi{}, Marko},
  journal = {Phys. Rev. E},
  volume = {92},
  issue = {4},
  pages = {042143},
  numpages = {17},
  year = {2015},
  month = {Oct},
  publisher = {American Physical Society},
  doi = {10.1103/PhysRevE.92.042143},
  url = {https://link.aps.org/doi/10.1103/PhysRevE.92.042143}
}

@article{ToddBrun2000,
  title = {Continuous measurements, quantum trajectories, and decoherent histories},
  author = {Brun, Todd A.},
  journal = {Phys. Rev. A},
  volume = {61},
  issue = {4},
  pages = {042107},
  numpages = {17},
  year = {2000},
  month = {Mar},
  publisher = {American Physical Society},
  doi = {10.1103/PhysRevA.61.042107},
  url = {https://link.aps.org/doi/10.1103/PhysRevA.61.042107}
}

@article{PeterZoller2010,
  title = {Dynamical Phase Transitions and Instabilities in Open Atomic Many-Body Systems},
  author = {Diehl, Sebastian and Tomadin, Andrea and Micheli, Andrea and Fazio, Rosario and Zoller, Peter},
  journal = {Phys. Rev. Lett.},
  volume = {105},
  issue = {1},
  pages = {015702},
  numpages = {4},
  year = {2010},
  month = {Jul},
  publisher = {American Physical Society},
  doi = {10.1103/PhysRevLett.105.015702},
  url = {https://link.aps.org/doi/10.1103/PhysRevLett.105.015702}
}

@article{Wang2023,
  title = {Accelerating relaxation dynamics in open quantum systems with Liouvillian skin effect},
  author = {Wang, Zeqing and Lu, Yao and Peng, Yi and Qi, Ran and Wang, Yucheng and Jie, Jianwen},
  journal = {Phys. Rev. B},
  volume = {108},
  issue = {5},
  pages = {054313},
  numpages = {8},
  year = {2023},
  month = {Aug},
  publisher = {American Physical Society},
  doi = {10.1103/PhysRevB.108.054313},
  url = {https://link.aps.org/doi/10.1103/PhysRevB.108.054313}
}

@article{ThomasBarthel2013,
  title = {Algebraic versus Exponential Decoherence in Dissipative Many-Particle Systems},
  author = {Cai, Zi and Barthel, Thomas},
  journal = {Phys. Rev. Lett.},
  volume = {111},
  issue = {15},
  pages = {150403},
  numpages = {5},
  year = {2013},
  month = {Oct},
  publisher = {American Physical Society},
  doi = {10.1103/PhysRevLett.111.150403},
  url = {https://link.aps.org/doi/10.1103/PhysRevLett.111.150403}
}

@article{solanki2025,
      title={Universal relaxation speedup in open quantum systems through transient conditional and unconditional resetting}, 
      author={Parvinder Solanki and Igor Lesanovsky and Gabriele Perfetto},
      year={2025},
      journal={arXiv:2512.10005},
      url={https://arxiv.org/abs/2512.10005}, 
}

@article{Kraus2008,
	author = {Diehl, S. and Micheli, A. and Kantian, A. and Kraus, B. and B{\"u}chler, H. P. and Zoller, P.},
	doi = {10.1038/nphys1073},
	journal = {Nature Physics},
	number = {11},
	pages = {878--883},
	title = {Quantum states and phases in driven open quantum systems with cold atoms},
	url = {https://doi.org/10.1038/nphys1073},
	volume = {4},
	year = {2008}}

@article{Igor2021,
  title = {Exponentially Accelerated Approach to Stationarity in Markovian Open Quantum Systems through the Mpemba Effect},
  author = {Carollo, Federico and Lasanta, Antonio and Lesanovsky, Igor},
  journal = {Phys. Rev. Lett.},
  volume = {127},
  issue = {6},
  pages = {060401},
  numpages = {6},
  year = {2021},
  month = {Aug},
  publisher = {American Physical Society},
  doi = {10.1103/PhysRevLett.127.060401},
  url = {https://link.aps.org/doi/10.1103/PhysRevLett.127.060401}
}

@article{Chatterjee2023,
  title = {Quantum Mpemba Effect in a Quantum Dot with Reservoirs},
  author = {Chatterjee, Amit Kumar and Takada, Satoshi and Hayakawa, Hisao},
  journal = {Phys. Rev. Lett.},
  volume = {131},
  issue = {8},
  pages = {080402},
  numpages = {6},
  year = {2023},
  month = {Aug},
  publisher = {American Physical Society},
  doi = {10.1103/PhysRevLett.131.080402},
  url = {https://link.aps.org/doi/10.1103/PhysRevLett.131.080402}
}

@article{Igor2022,
  title = {Accelerating the approach of dissipative quantum spin systems towards stationarity through global spin rotations},
  author = {Kochsiek, Simon and Carollo, Federico and Lesanovsky, Igor},
  journal = {Phys. Rev. A},
  volume = {106},
  issue = {1},
  pages = {012207},
  numpages = {7},
  year = {2022},
  month = {Jul},
  publisher = {American Physical Society},
  doi = {10.1103/PhysRevA.106.012207},
  url = {https://link.aps.org/doi/10.1103/PhysRevA.106.012207}
}

@article{Ranjan2024,
  title = {Quest for optimal quantum resetting: Protocols for a particle on a chain},
  author = {Chatterjee, Pallabi and Aravinda, S. and Modak, Ranjan},
  journal = {Phys. Rev. E},
  volume = {110},
  issue = {3},
  pages = {034132},
  numpages = {14},
  year = {2024},
  month = {Sep},
  publisher = {American Physical Society},
  doi = {10.1103/PhysRevE.110.034132},
  url = {https://link.aps.org/doi/10.1103/PhysRevE.110.034132}
}

@article{Ljubotina2017, 
	author = {Ljubotina, Marko and {\v Z}nidari{\v c}, Marko and Prosen, Toma{\v z}},
	journal = {Nature Communications},
	title = {Spin diffusion from an inhomogeneous quench in an integrable system},
	url = {https://doi.org/10.1038/ncomms16117},
	volume = {8},
	year = {2017}}

@article{beato2025,
      title={Relaxation control of open quantum systems}, 
      author={Nicolò Beato and Gianluca Teza},
      year={2025},
      journal={arXiv:2507.15948},
      url={https://arxiv.org/abs/2507.15948}, 
}

@article{Torres-Herrera_2014,
doi = {10.1088/1367-2630/16/6/063010},
url = {https://doi.org/10.1088/1367-2630/16/6/063010},
year = {2014},
month = {jun},
publisher = {IOP Publishing},
volume = {16},
number = {6},
pages = {063010},
author = {Torres-Herrera, E J and Vyas, Manan and Santos, Lea F},
title = {General features of the relaxation dynamics of interacting quantum systems},
journal = {New Journal of Physics}
}

@article{Dante2025,
  title = {Relaxation dynamics of a quantum spin coupled to a topological edge state},
  author = {Liu, Qiyu and Karrasch, Christoph and Kennes, Dante Marvin and Rausch, Roman},
  journal = {Phys. Rev. B},
  volume = {111},
  issue = {23},
  pages = {235153},
  numpages = {9},
  year = {2025},
  month = {Jun},
  publisher = {American Physical Society},
  doi = {10.1103/5qpv-b7l3},
  url = {https://link.aps.org/doi/10.1103/5qpv-b7l3}
}

@article{Adam2024, 
	author = {Wu, Ling-Na and Nettersheim, Jens and Fe{\ss}, Julian and Schnell, Alexander and Burgardt, Sabrina and Hiebel, Silvia and Adam, Daniel and Eckardt, Andr{\'e} and Widera, Artur},
	doi = {10.1038/s41467-024-46054-9},
	journal = {Nature Communications},
	number = {1},
	pages = {1714},
	title = {Indication of critical scaling in time during the relaxation of an open quantum system},
	url = {https://doi.org/10.1038/s41467-024-46054-9},
	volume = {15},
	year = {2024}}

@article{Zhang2020,
  title = {Anomalous relaxation and multiple timescales in the quantum XY model with boundary dissipation},
  author = {Zhang, Shun-Yao and Gong, Ming and Guo, Guang-Can and Zhou, Zheng-Wei},
  journal = {Phys. Rev. B},
  volume = {101},
  issue = {15},
  pages = {155150},
  numpages = {13},
  year = {2020},
  month = {Apr},
  publisher = {American Physical Society},
  doi = {10.1103/PhysRevB.101.155150},
  url = {https://link.aps.org/doi/10.1103/PhysRevB.101.155150}
}

@article{Caspel2018,
  title = {Symmetry-protected coherent relaxation of open quantum systems},
  author = {van Caspel, Moos and Gritsev, Vladimir},
  journal = {Phys. Rev. A},
  volume = {97},
  issue = {5},
  pages = {052106},
  numpages = {8},
  year = {2018},
  month = {May},
  publisher = {American Physical Society},
  doi = {10.1103/PhysRevA.97.052106},
  url = {https://link.aps.org/doi/10.1103/PhysRevA.97.052106}
}

@article{Tamir2013,
    author = {Tamir, Boaz and Cohen, Eliahu},
    year = {2013},
    month = {05},
    pages = {7-17},
    title = {Introduction to Weak Measurements and Weak Values},
    volume = {2},
    journal = {QUANTA},
    doi = {10.12743/quanta.v2i1.14}
}

@article{Bao2025,
  title = {Accelerating Quantum Relaxation via Temporary Reset: A Mpemba-Inspired Approach},
  author = {Bao, Ruicheng and Hou, Zhonghuai},
  journal = {Phys. Rev. Lett.},
  volume = {135},
  issue = {15},
  pages = {150403},
  numpages = {10},
  year = {2025},
  month = {Oct},
  publisher = {American Physical Society},
  doi = {10.1103/g94p-7421},
  url = {https://link.aps.org/doi/10.1103/g94p-7421}
}

@article{Sarang2022,
author = {Jacopo De Nardis  and Sarang Gopalakrishnan  and Romain Vasseur  and Brayden Ware },
title = {Subdiffusive hydrodynamics of nearly integrable anisotropic spin chains},
journal = {Proceedings of the National Academy of Sciences},
volume = {119},
number = {34},
year = {2022},
doi = {10.1073/pnas.2202823119},
URL = {https://www.pnas.org/doi/abs/10.1073/pnas.2202823119}}

@article{Andrea2024,
  title = {Mpemba Effects in Open Nonequilibrium Quantum Systems},
  author = {Nava, Andrea and Egger, Reinhold},
  journal = {Phys. Rev. Lett.},
  volume = {133},
  issue = {13},
  pages = {136302},
  numpages = {7},
  year = {2024},
  month = {Sep},
  publisher = {American Physical Society},
  doi = {10.1103/PhysRevLett.133.136302},
  url = {https://link.aps.org/doi/10.1103/PhysRevLett.133.136302}
}

@article{liu2025,
      title={A General Strategy for Realizing Mpemba Effects in Open Quantum Systems}, 
      author={Yaru Liu and Yucheng Wang},
      year={2025},
      journal={arXiv:2511.04354},
      url={https://arxiv.org/abs/2511.04354}, 
}

@article{Zhang_Mpemba2025,
	author = {Zhang, Jie and Xia, Gang and Wu, Chun-Wang and Chen, Ting and Zhang, Qian and Xie, Yi and Su, Wen-Bo and Wu, Wei and Qiu, Cheng-Wei and Chen, Ping-Xing and Li, Weibin and Jing, Hui and Zhou, Yan-Li},
	doi = {10.1038/s41467-024-54303-0},
	id = {Zhang2025},
	isbn = {2041-1723},
	journal = {Nature Communications},
	number = {1},
	pages = {301},
	title = {Observation of quantum strong Mpemba effect},
	url = {https://doi.org/10.1038/s41467-024-54303-0},
	volume = {16},
	year = {2025}}

@article{caldas2025,
      title={Exponentially accelerated relaxation and quantum Mpemba effect in open quantum systems}, 
      author={Emerson Lima Caldas and Diego Paiva Pires},
      year={2025},
      journal={arXiv:2512.07561},
      url={https://arxiv.org/abs/2512.07561}, 
}

@article{Archak2025,
  title = {Non-Markovian Quantum Mpemba Effect},
  author = {Strachan, David J. and Purkayastha, Archak and Clark, Stephen R.},
  journal = {Phys. Rev. Lett.},
  volume = {134},
  issue = {22},
  pages = {220403},
  numpages = {7},
  year = {2025},
  month = {Jun},
  publisher = {American Physical Society},
  doi = {10.1103/PhysRevLett.134.220403},
  url = {https://link.aps.org/doi/10.1103/PhysRevLett.134.220403}
}

@article{longhi2025mpemba,
  title={Mpemba effect and super-accelerated thermalization in the damped quantum harmonic oscillator},
  author={Longhi, Stefano},
  journal={Quantum},
  volume={9},
  pages={1677},
  year={2025},
  url={https://quantum-journal.org/papers/q-2025-03-26-1677/},
}

@article{bagui2025,
      title={Detection of Mpemba effect through good observables in open quantum systems}, 
      author={Pitambar Bagui and Arijit Chatterjee and Bijay Kumar Agarwalla},
      year={2025},
      journal={arXiv:2512.02709},
      url={https://arxiv.org/abs/2512.02709}, 
}

@article{Xhek2025,
  title = {Quantum Mpemba Effect in Random Circuits},
  author = {Turkeshi, Xhek and Calabrese, Pasquale and De Luca, Andrea},
  journal = {Phys. Rev. Lett.},
  volume = {135},
  issue = {4},
  pages = {040403},
  numpages = {10},
  year = {2025},
  month = {Jul},
  publisher = {American Physical Society},
  doi = {10.1103/5d6p-8d1b},
  url = {https://link.aps.org/doi/10.1103/5d6p-8d1b}
}

@article{di2025,
  title={Measurement-induced symmetry restoration and quantum mpemba effect},
  author={Di Giulio, Giuseppe and Turkeshi, Xhek and Murciano, Sara},
  journal={Entropy},
  volume={27},
  number={4},
  pages={407},
  year={2025},
  publisher={MDPI},
  url={https://www.mdpi.com/1099-4300/27/4/407}
}

@article{Yu2025,
	author = {Yu, Hui and Liu, Shuo and Zhang, Shi-Xin},
	journal = {AAPPS Bulletin},
	number = {1},
	pages = {17},
	title = {Quantum Mpemba effects from symmetry perspectives},
	url = {https://doi.org/10.1007/s43673-025-00157-7},
	volume = {35},
	year = {2025}}

@article{Liu2024,
  title = {Symmetry Restoration and Quantum Mpemba Effect in Symmetric Random Circuits},
  author = {Liu, Shuo and Zhang, Hao-Kai and Yin, Shuai and Zhang, Shi-Xin},
  journal = {Phys. Rev. Lett.},
  volume = {133},
  issue = {14},
  pages = {140405},
  numpages = {7},
  year = {2024},
  month = {Oct},
  publisher = {American Physical Society},
  doi = {10.1103/PhysRevLett.133.140405},
  url = {https://link.aps.org/doi/10.1103/PhysRevLett.133.140405}
}

@article{Yamashika2024,
  title = {Entanglement asymmetry and quantum Mpemba effect in two-dimensional free-fermion systems},
  author = {Yamashika, Shion and Ares, Filiberto and Calabrese, Pasquale},
  journal = {Phys. Rev. B},
  volume = {110},
  issue = {8},
  pages = {085126},
  numpages = {16},
  year = {2024},
  month = {Aug},
  publisher = {American Physical Society},
  doi = {10.1103/PhysRevB.110.085126},
  url = {https://link.aps.org/doi/10.1103/PhysRevB.110.085126}
}

@article{ulcakar2025,
      title={Conserved quantities enable the quantum Mpemba effect in weakly open systems}, 
      author={Iris Ulčakar and Rustem Sharipov and Gianluca Lagnese and Zala Lenarčič},
      year={2025},
      journal={arXiv:2511.16739},
      url={https://arxiv.org/abs/2511.16739}, 
}

@article{Tanmay2025,
  title = {Quantum Mpemba effect without global symmetries},
  author = {Bhore, Tanmay and Su, Lei and Martin, Ivar and Clerk, Aashish A. and Papi\ifmmode \acute{c}\else \'{c}\fi{}, Zlatko},
  journal = {Phys. Rev. B},
  volume = {112},
  issue = {12},
  pages = {L121109},
  numpages = {6},
  year = {2025},
  month = {Sep},
  publisher = {American Physical Society},
  doi = {10.1103/1td3-2vwf},
  url = {https://link.aps.org/doi/10.1103/1td3-2vwf}
}

@article{Filiberto2024,
  title = {Microscopic Origin of the Quantum Mpemba Effect in Integrable Systems},
  author = {Rylands, Colin and Klobas, Katja and Ares, Filiberto and Calabrese, Pasquale and Murciano, Sara and Bertini, Bruno},
  journal = {Phys. Rev. Lett.},
  volume = {133},
  issue = {1},
  pages = {010401},
  numpages = {6},
  year = {2024},
  month = {Jul},
  publisher = {American Physical Society},
  doi = {10.1103/PhysRevLett.133.010401},
  url = {https://link.aps.org/doi/10.1103/PhysRevLett.133.010401}
}

@article{Ares2025,
	author = {Ares, Filiberto and Calabrese, Pasquale and Murciano, Sara},
	journal = {Nature Reviews Physics},
	number = {8},
	title = {The quantum Mpemba effects},
	url = {https://doi.org/10.1038/s42254-025-00838-0},
	volume = {7},
	year = {2025}}

@article{Schiro_2024,
  title = {Many-Body Dynamics in Monitored Atomic Gases without Postselection Barrier},
  author = {Passarelli, Gianluca and Turkeshi, Xhek and Russomanno, Angelo and Lucignano, Procolo and Schir\`o, Marco and Fazio, Rosario},
  journal = {Phys. Rev. Lett.},
  volume = {132},
  issue = {16},
  pages = {163401},
  numpages = {7},
  year = {2024},
  month = {Apr},
  publisher = {American Physical Society},
  doi = {10.1103/PhysRevLett.132.163401},
  url = {https://link.aps.org/doi/10.1103/PhysRevLett.132.163401}
}

@article{Marco2024,
  title = {Entanglement Dynamics in Monitored Systems and the Role of Quantum Jumps},
  author = {Le Gal, Youenn and Turkeshi, Xhek and Schir\`o, Marco},
  journal = {PRX Quantum},
  volume = {5},
  issue = {3},
  pages = {030329},
  numpages = {21},
  year = {2024},
  month = {Aug},
  publisher = {American Physical Society},
  doi = {10.1103/PRXQuantum.5.030329},
  url = {https://link.aps.org/doi/10.1103/PRXQuantum.5.030329}
}

@article{PeterZoller2018,
  title = {Theory of a Quantum Scanning Microscope for Cold Atoms},
  author = {Yang, D. and Laflamme, C. and Vasilyev, D. V. and Baranov, M. A. and Zoller, P.},
  journal = {Phys. Rev. Lett.},
  volume = {120},
  issue = {13},
  pages = {133601},
  numpages = {6},
  year = {2018},
  month = {Mar},
  publisher = {American Physical Society},
  doi = {10.1103/PhysRevLett.120.133601},
  url = {https://link.aps.org/doi/10.1103/PhysRevLett.120.133601}
}

@article{Yuen_83,
author = {Horace P. Yuen and Vincent W. S. Chan},
journal = {Opt. Lett.},
number = {3},
pages = {177--179},
title = {Noise in homodyne and heterodyne detection},
volume = {8},
year = {1983},
url = {https://opg.optica.org/ol/abstract.cfm?URI=ol-8-3-177},
doi = {10.1364/OL.8.000177}}

@article{Warszawski_2002,
   title={Quantum trajectories for realistic photodetection: I. General formalism},
   volume={5},
   url={http://dx.doi.org/10.1088/1464-4266/5/1/301},
   number={1},
   journal={Journal of Optics B: Quantum and Semiclassical Optics},
   author={Warszawski, P and Wiseman, H M},
   year={2002}}

@article{Clerk2025,
  title = {Accelerating Dissipative State Preparation with Adaptive Open Quantum Dynamics},
  author = {Pocklington, Andrew and Clerk, Aashish A.},
  journal = {Phys. Rev. Lett.},
  volume = {134},
  issue = {5},
  pages = {050603},
  numpages = {7},
  year = {2025},
  month = {Feb},
  publisher = {American Physical Society},
  doi = {10.1103/PhysRevLett.134.050603},
  url = {https://link.aps.org/doi/10.1103/PhysRevLett.134.050603}
}

@article{Tuorila2017,
	author = {Tuorila, Jani and Partanen, Matti and Ala-Nissila, Tapio and M{\"o}tt{\"o}nen, Mikko},
	journal = {npj Quantum Information},
	number = {1},
	title = {Efficient protocol for qubit initialization with a tunable environment},
	url = {https://doi.org/10.1038/s41534-017-0027-1},
	volume = {3},
	year = {2017}}

@article{Liu_2025,
  title = {Optimally Fast Qubit Reset},
  author = {Liu, Yue and Huang, Chenlong and Zhang, Xingyu and He, Dahai},
  journal = {Phys. Rev. Lett.},
  volume = {134},
  issue = {10},
  pages = {100401},
  numpages = {6},
  year = {2025},
  month = {Mar},
  publisher = {American Physical Society},
  doi = {10.1103/PhysRevLett.134.100401},
  url = {https://link.aps.org/doi/10.1103/PhysRevLett.134.100401}
}

@article{Zoller1992,
  title = {Monte Carlo simulation of the atomic master equation for spontaneous emission},
  author = {Dum, R. and Zoller, P. and Ritsch, H.},
  journal = {Phys. Rev. A},
  volume = {45},
  issue = {7},
  pages = {4879--4887},
  numpages = {0},
  year = {1992},
  month = {Apr},
  publisher = {American Physical Society},
  doi = {10.1103/PhysRevA.45.4879},
  url = {https://link.aps.org/doi/10.1103/PhysRevA.45.4879}
}

@article{Molmer_93,
author = {Klaus M{\o}lmer and Yvan Castin and Jean Dalibard},
journal = {J. Opt. Soc. Am. B},
number = {3},
title = {Monte Carlo wave-function method in quantum optics},
volume = {10},
year = {1993},
url = {https://opg.optica.org/josab/abstract.cfm?URI=josab-10-3-524}
}

@article{vasseur2025,
      title={Mixed-state learnability transitions in monitored noisy quantum dynamics}, 
      author={Hansveer Singh and Romain Vasseur and Andrew C. Potter and Sarang Gopalakrishnan},
      year={2025},
      journal={arXiv:2503.10308},
      url={https://arxiv.org/abs/2503.10308}, 
}

@article{Calabrese2023,
	author = {Ares, Filiberto and Murciano, Sara and Calabrese, Pasquale},
	journal = {Nature Communications},
	number = {1},
	title = {Entanglement asymmetry as a probe of symmetry breaking},
	url = {https://doi.org/10.1038/s41467-023-37747-8},
	volume = {14},
	year = {2023}}

@article{Heng2025,
  title = {Tuning the quantum Mpemba effect in an isolated system by initial-state engineering},
  author = {Yu, Yi-Han and Jin, Tian-Ren and Zhang, Lv and Xu, Kai and Fan, Heng},
  journal = {Phys. Rev. B},
  volume = {112},
  issue = {9},
  pages = {094315},
  numpages = {12},
  year = {2025},
  month = {Sep},
  publisher = {American Physical Society},
  doi = {10.1103/yzjd-pk8h},
  url = {https://link.aps.org/doi/10.1103/yzjd-pk8h}
}

@article{Ganguly_2025,
doi = {10.1088/1742-5468/ae1ffa},
url = {https://doi.org/10.1088/1742-5468/ae1ffa},
year = {2025},
month = {dec},
publisher = {IOP Publishing},
volume = {2025},
number = {12},
pages = {123102},
author = {Ganguly, Katha and Gopalakrishnan, Preethi and Naik, Atharva and Kumar Agarwalla, Bijay and Kulkarni, Manas},
title = {Quantum trajectories and Page-curve entanglement dynamics},
journal = {Journal of Statistical Mechanics: Theory and Experiment}
}

@article{bagui_chatterjee2025,
      title={Accelerated relaxation and Mpemba-like effect for operators in open quantum systems}, 
      author={Pitambar Bagui and Arijit Chatterjee and Bijay Kumar Agarwalla},
      year={2025},
      journal={arXiv:2510.24630},
      url={https://arxiv.org/abs/2510.24630} 
}

@article{YanBin2026,
  title = {Deterministic Quantum Trajectory via Imaginary Time Evolution},
  author = {Mittal, Shivan and Yan, Bin},
  journal = {Phys. Rev. Lett.},
  volume = {136},
  issue = {1},
  pages = {010401},
  numpages = {6},
  year = {2026},
  month = {Jan},
  publisher = {American Physical Society},
  doi = {10.1103/fgkg-n2b9},
  url = {https://link.aps.org/doi/10.1103/fgkg-n2b9}
}

\begin{center}
\section*{End Matter}
\end{center}
\section*{Continuous monitoring of total magnetization $\hat{S}_z$ -- Lack of symmetry restoration in Quantum Jump}
\begin{figure}
    \centering
    \includegraphics[width=0.5\linewidth]{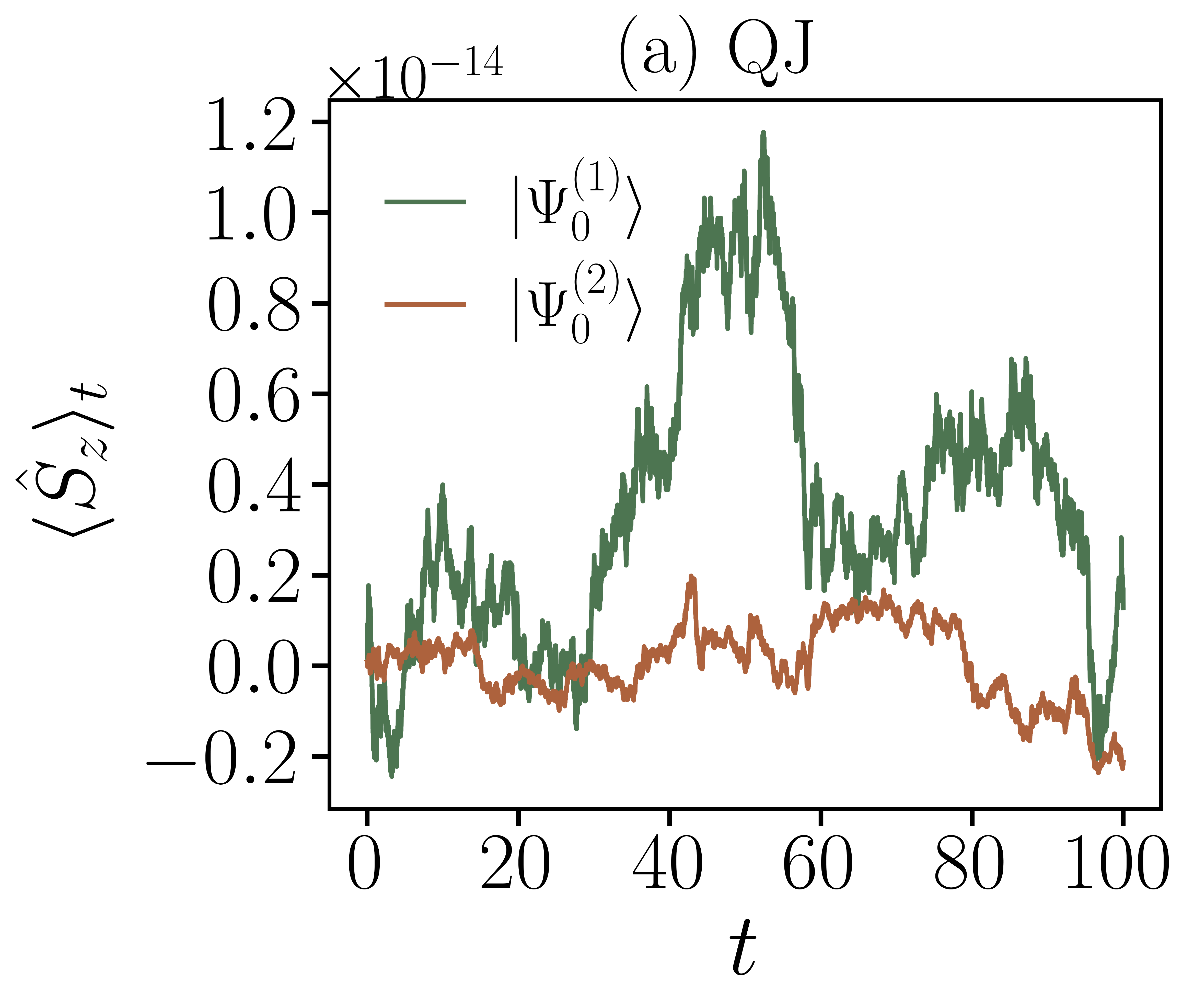}%
    \includegraphics[width=0.5\linewidth]{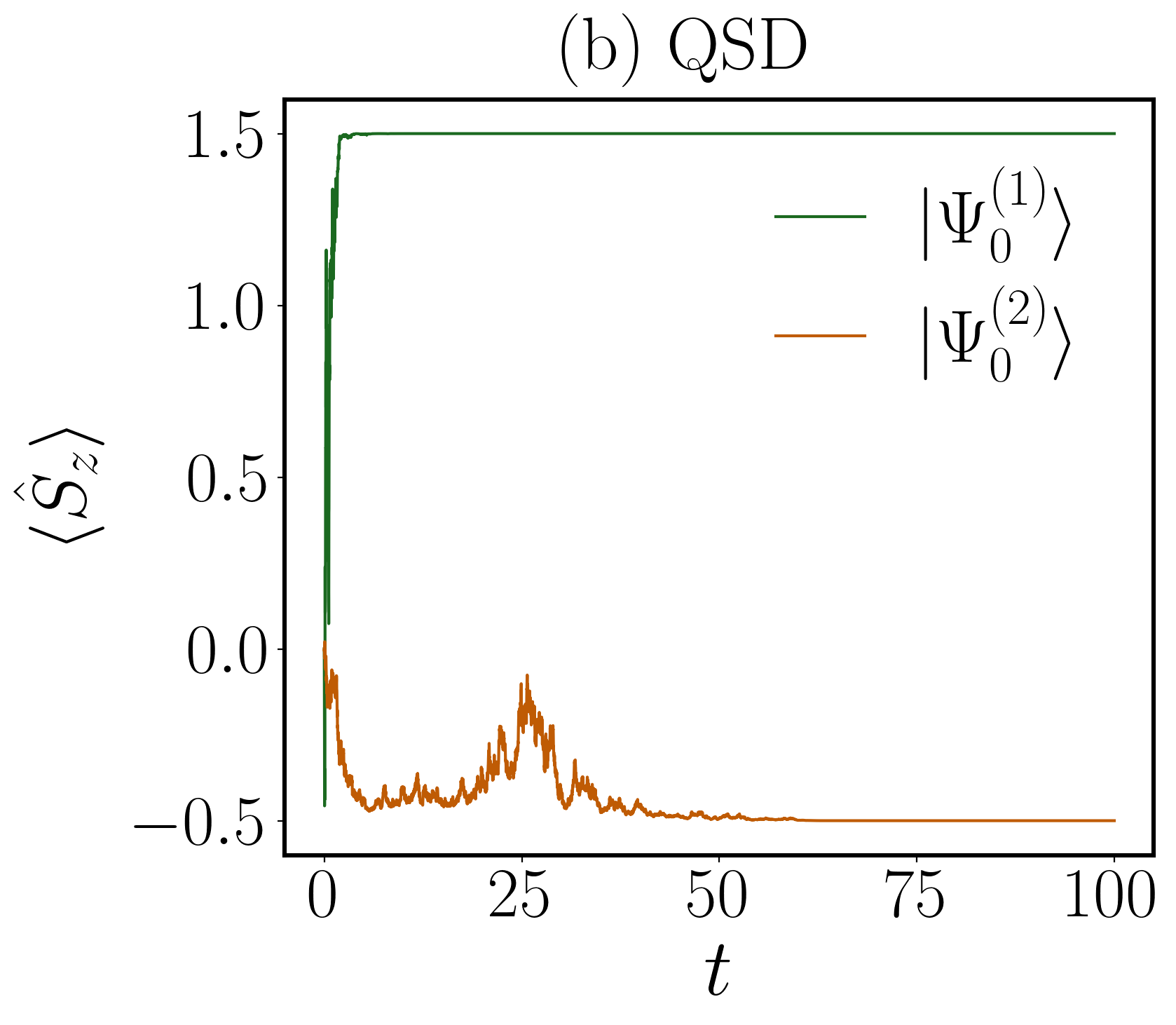}
    \caption{Continuous monitoring of the total magnetization operator $\hat{S}_z$ under (a) quantum jump (QJ) process and (b) quantum state diffusion (QSD) process. The evolution is initialized from a non-$U(1)$ preserving initial state of the form $|\Psi_S\ra=\big(|n-L/2\ra+|-n+L/2\ra\big)/\sqrt{2}$ where $|n-L/2\ra$ corresponds to an eigenstate of the operator $\hat{S}_z$ with eigenvalue $(n-L/2)$ which contains $n$ number of up spins and $L-n$ number of down spins. Under continuous measurements combined with unitary evolution by $U(1)$ conserving $H_S$ (XX Hamiltonian with $J=1$ and $\Delta=0.1$), the quantum jump process is inefficient to restore the symmetry in (a). However, under QSD in (b), the $U(1)$ symmetry is restored and a seperation of time-scale is observed. The other parameters are $L=5$, $\gamma=0.1$, $dt=0.01$.}
    \label{fig:QJ_No_SR}
\end{figure}

We discuss here the lack of symmetry restoration under the continuous monitoring of the total magnetization operator $\hat{S}_z$ in the quantum jump processes. The root cause of this problem is that $\hat{S}_z$ has a symmetric eigen-spectra i.e, it always has a $-\lambda$ eigenvalue corresponding to a $+\lambda$ eigenvalue. We show that starting from an initial state $|\Psi_S\ra=\alpha|+\lambda\ra+\beta|-\lambda\ra$, the system never restores the symmetry under quantum jump monitoring of $\hat{S}_z$. Here, all the ancilla are considered to be spin-1/2 particles, prepared in $|0\ra_A$ state. At each time step $dt$, one such ancilla is coupled to the system operator $\hat{S}_z$ through the unitary, $U(dt)=e^{-i\sqrt{\gamma dt}\hat{S}_z\otimes \hat{\sigma}_y}$. Therefore, the joint state of the system-ancilla up to $O(dt)$ is,
\begin{equation}
    |\Psi_{SA}\ra=|\Psi_S\ra|0\ra_A\!-\!\frac{\gamma dt}{2}\hat{S}_z^2|\Psi_S\ra|0\ra_A\!+\!\sqrt{\gamma dt}\hat{S}_z|\Psi_S\ra |1\ra_A
\end{equation}
A projective measurement of the ancilla operator $\hat{\sigma}_z$ conditions the system evolution as the following: if the outcome is $0$ and $1$, the post-measurement state of the system following Eq.~\eqref{eq:P-M_state} is,
\begin{align}
    &|\Psi_{S,0}(dt)\ra=\frac{\Big[\alpha \,d_0(\lambda)|+\lambda\ra+\beta\, d_0(-\lambda)|-\lambda\ra\Big]}{|\alpha|^2|d_0(\lambda)|^2+|\beta|^2|d_0(-\lambda)|^2},\nonumber\\
    &|\Psi_{S,1}(dt)\ra=\frac{\Big[\alpha \,d_1(\lambda)|+\lambda\ra+\beta\, d_1(-\lambda)|-\lambda\ra\Big]}{|\alpha|^2|d_0(\lambda)|^2+|\beta|^2|d_0(-\lambda)|^2},
\end{align}
where $d_0(\lambda)=d_0(-\lambda)=\cos(\lambda \sqrt{\gamma dt})$ and $d_1(\lambda)=\sin(\lambda \sqrt{\gamma dt})$, $d_1(-\lambda)=-\sin(\lambda \sqrt{\gamma dt})$. Hence, under QJ monitoring $d_0(\lambda)$ and $d_0(-\lambda)$ can never be distinguished, consequence of which leads to lack of symmetry restoration, as presented in Fig.~\ref{fig:QJ_No_SR}(a).

However, we will now show that under QSD protocol, $U(1)$ symmetry can be restored even under $\hat{S}_z$ measurement. Here, all the ancilla are considered to be quantum harmonic oscillators, prepared in the ground state $\Psi_A(x)=(2\gamma dt/\pi)^{1/4}e^{-\gamma dt x^2}$. At each time step $dt$, one such ancilla is coupled to the system operator $\hat{S}_z$ through the unitary, $U(dt)=e^{-i \hat{S}_z\otimes\hat{P}_A}$ where $\hat{P}_A$ is the momentum operator of the ancilla. Therefore, the joint system-ancilla state is,
\begin{align}
    |\Psi_{SA}\ra=\Bigg(\frac{2\gamma dt}{\pi}\Bigg)^{1/4}\int_{-\infty}^{\infty} dx_A\,\Big[ \alpha e^{-\gamma dt(x-\lambda)^2}|\lambda\ra|x_A\ra \nonumber\\+ \beta e^{-\gamma dt (x+\lambda)^2}|-\lambda\ra|x_A\ra\Big]
\end{align}
Under the projective measurement of the position of the ancilla, the conditional state of the system becomes,
\begin{align}
    |\Psi_{S,x_0}(dt)\ra=\frac{\alpha\, d_{x_0}(\lambda)|\lambda\ra+\beta \,d_{x_0}(-\lambda) |-\lambda\ra}{|\alpha|^2|d_{x_0}(\lambda)|^2+|\beta|^2|d_{x_0}(-\lambda)|^2},
\end{align}
where $d_{x_0}(\pm\lambda)=(2\gamma dt/\pi)^{1/4}e^{-\gamma dt(x_0\pm\lambda)^2}$, yielding distinct contributions for $+ \lambda$ and $-\lambda$. Need further explanation after this expression. Hence, under QSD, the state of the full system can relax to the eigenstate of $\hat{S}_z$ and thus restores $U(1)$ symmetry which can be observed in Fig.~\ref{fig:QJ_No_SR}(b).

\vspace{0.2cm}

\section*{Absence of measurement backaction: classical noise protocol}
\begin{figure}
    \centering
    \includegraphics[width=0.5\linewidth]{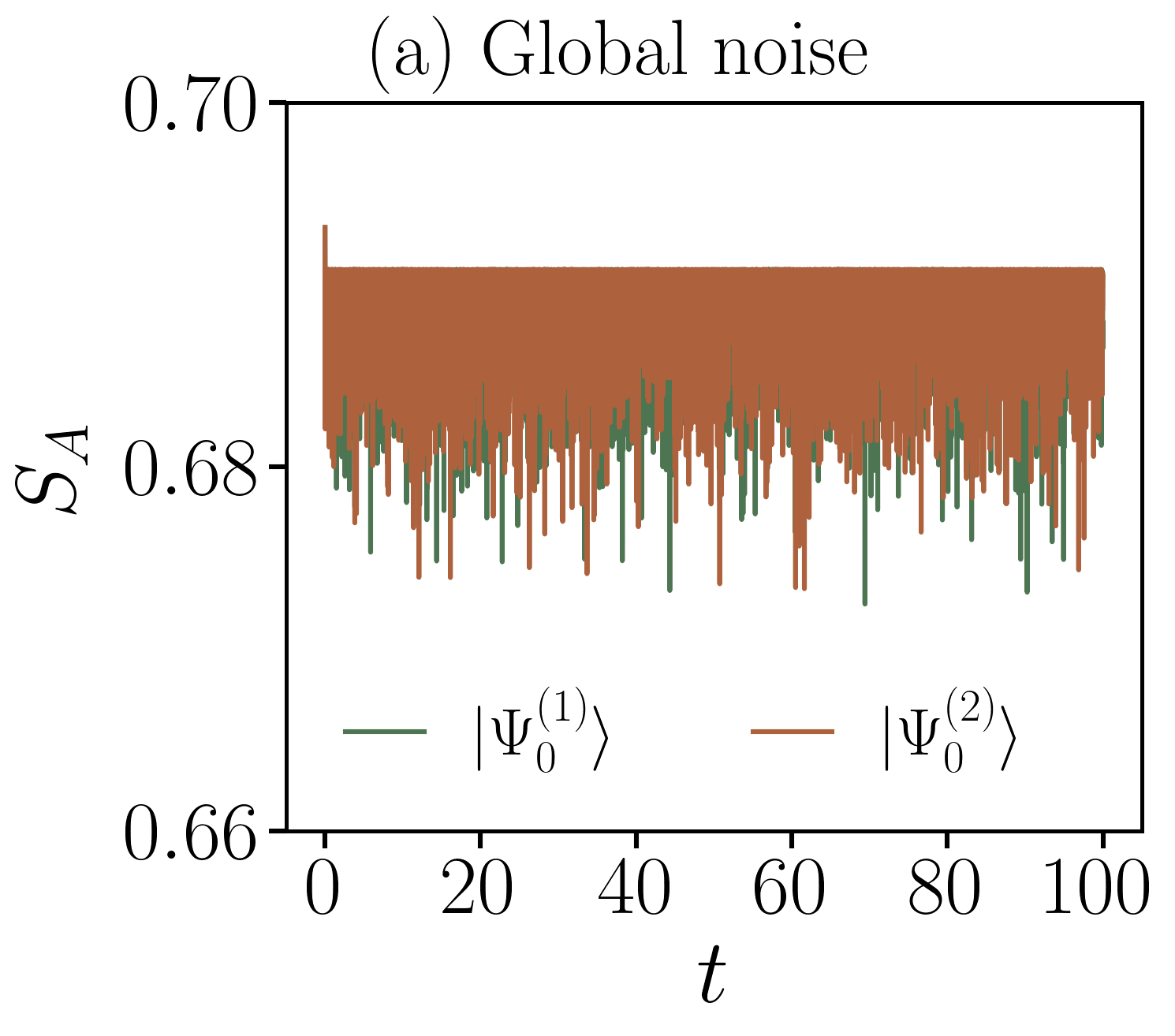}%
    \includegraphics[width=0.5\linewidth]{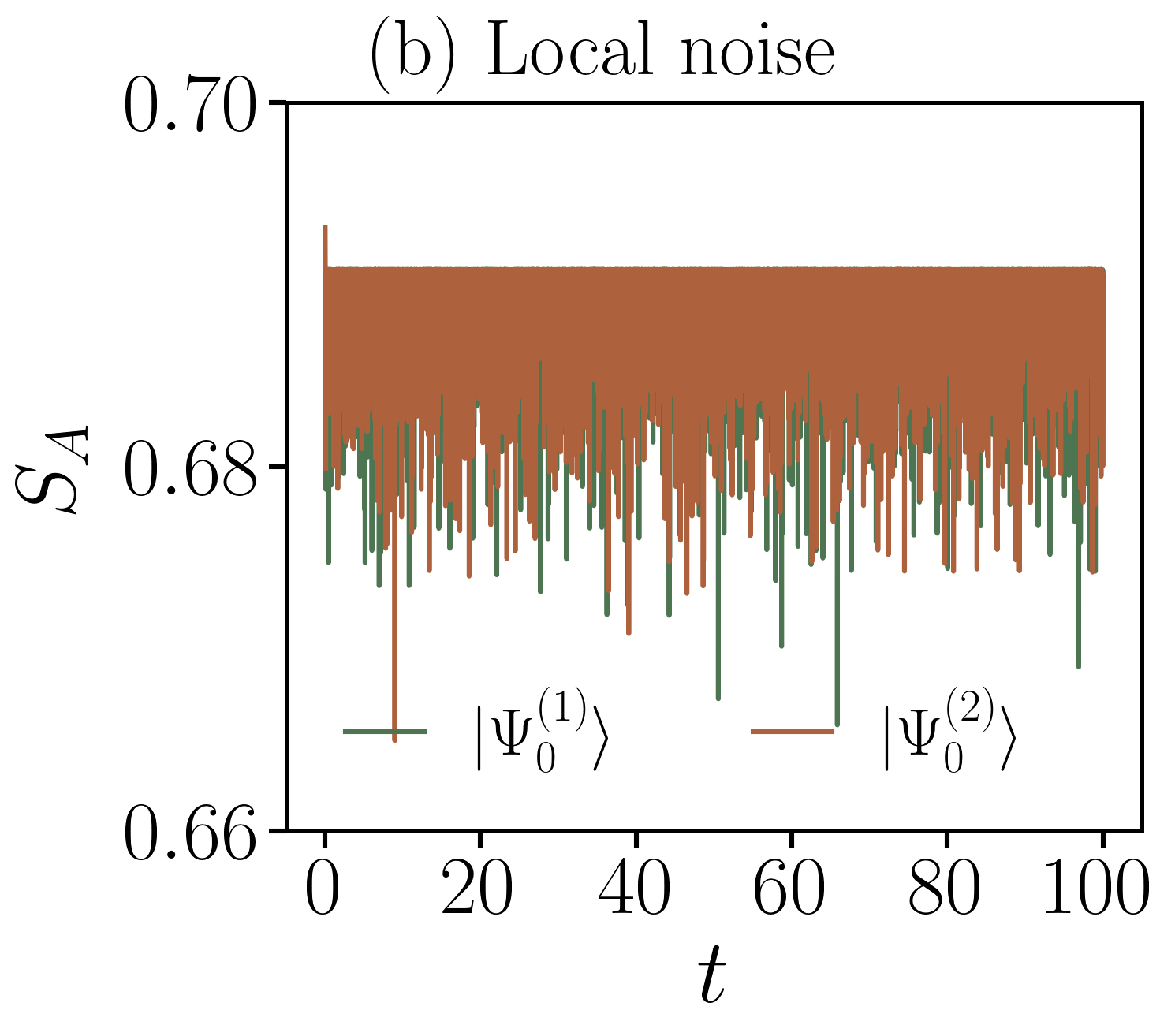}
    \caption{Plot of entropy asymmetry $S_A$ defined in Eq.~\eqref{eq:entropy_asym} under classical noise protocol: (a) A global noise $\xi(t)$ is coupled to total number operator $\hat{N}$ of the system. (b) Independent local noise $\xi_i(t)$ is coupled to local number operator $\hat{n}_i$ throughout the lattice. There is no symmetry restoration in both cases.}
    \label{fig:ent_asym}
\end{figure}

It is important to note that the symmetry restoration effect, as seen in the main text, emerges solely due to the back-action of quantum measurement and is completely absent when a quantum trajectory protocol does not mimic a quantum measurement. For example, one can devise a protocol, where the system couples to a global classical noise,  $\hat{H}_\xi=\hat{H}+\xi(t)\hat{N}$, where $\xi(t)$ is a Gaussian white noise with $\overline{\xi(t)}=0$ and $\overline{\xi(t) \xi(t')}=\gamma\,\delta(t-t')$. Note that for this protocol, ensemble average of quantum trajectories results to the same GKSL master equation as of global monitoring,  i.e., $\partial\rho/\partial t =-i[\hat{H},\rho]+\gamma\,\big[\hat{N}\rho  \hat{N}-\frac{1}{2}\{\hat{N}^2,\rho\}\big]$ with jump operators being $\hat{N}$. Similar to global noise, one can also consider local noise throughout the lattice as $\hat{H}_\xi=\hat{H}+\sum_{i=1}^{L}\xi_i(t)\hat{n}_i$.
The ensemble average dynamics of local noise protocol is also same with local monitoring which is $\partial\rho/\partial t =-i[\hat{H},\rho]+\gamma\,\sum_{i=1}^{L}\big[\hat{n}_i\rho  \hat{n}_i-\frac{1}{2}\{\hat{n}_i,\rho\}\big]$. Note that $\hat{n}_i^2= \hat{n}_i$. Different realizations of the classical noise results different quantum trajectories $|\Psi_\xi(t)\ra$.
However, both global and local noise protocol do not mimic any quantum measurement and hence do not produce any backaction. In this case, irrespective of the initial state $d\la \hat{N}\ra_t=0$ for every trajectory. However, one can calculate the entropy asymmetry $S_A$ at each trajectory, defined as~\cite{Calabrese2023},
\begin{equation}
    S_A = -{\rm Tr}[\rho_{\rm sym}\ln \rho_{\rm sym}]+{\rm Tr}[\rho \ln \rho],\label{eq:entropy_asym}
\end{equation}
where $\rho=|\Psi_\xi(t)\ra\la\Psi_\xi(t)|$ and $\rho_{\rm sym}=\int_{-\infty}^{\infty} \frac{d\alpha}{2\pi}e^{-i\alpha \hat{N}}\,\rho\, e^{i\alpha\hat{N}}$ and it does not approaches zero under the classical noise protocol which implies no symmetry restoration. This is illustrated in Fig.~\ref{fig:ent_asym}(a) and (b) for global as well as local monitoring, respectively.

\newpage
\onecolumngrid

\setcounter{figure}{0}
\renewcommand{\thefigure}{S\arabic{figure}}
\setcounter{equation}{0}
\renewcommand{\theequation}{S\arabic{equation}}

\begin{center}
\textbf{Supplemental Material for ``Measurement induced faster symmetry restoration in quantum trajectories"}
\end{center}
\section{Alternate derivation of the equation of probability distribution $P(n,t)$ under global monitoring via QSD protocol}

In the main text, we have derived the equation for the probability distribution $P(n,t)=|c_n^{t}|^2$ under global monitoring via the QSD protocol, following the stochastic Schr\"odinger equation. In this supplementary, we will give an alternative derivation by using cumulant generating function and deriving the equation of cumulants.
Recall that the stochastic Schr\"odinger equation (SSE) for the QSD protocol describing global monitoring is
\begin{equation}
    d|\Psi_t\rangle\! =\!\Big[\!\! -i\hat{H}_S dt+(\hat{N}\!-\!\langle\hat{N}\rangle_t)d\xi_t\!-\!\frac{\gamma dt}{2}(\hat{N}\!-\!\langle\hat{N}\rangle_t)^2\Big]|\Psi_t\ra. \label{eq:SSE_QSD_supp}
\end{equation}
Following the Eq.~\eqref{eq:SSE_QSD_supp}, we have obtain the equation of $\la\hat{N}\ra$ in main text which is $d\la \hat{N}\ra_t=2 d\xi_t \Big(\la \hat{N}^2\ra_t-\la \hat{N}\ra^2_t\Big)$. Now following Eq.~\eqref{eq:SSE_QSD_supp}, we will obtain the equation of number fluctuation (second cumulant) as well as higher cumulants in quantum trajectories,
\begin{align}
    d\la \hat{N}^2\ra^c_t = 2 d\xi_t \la\hat{N}^3\ra_t^c, \quad d\la \hat{N}^k\ra_t^c = 2 d\xi_t \la\hat{N}^{k+1}\ra^c_t
\end{align}
where the superscript $c$ stands for cumulants. 
Once obtained the evolution equation for the cumulants, we can write the equation of the cumulant generating function $\mathcal{K}_t(\lambda)=\sum_{k=1}^{\infty}(i\lambda)^k \la\hat{N}^k\ra^c_t/k!$ as the following,
\begin{align}
    d\mathcal{K}_t(\lambda)=\sum_{k=1}^{\infty} \frac{(i\lambda)^k}{k!}d\la\hat{N}^k\ra^c_t=2d\xi_t\sum_{k=1}^{\infty}\frac{(i\lambda)^k}{k!} \la\hat{N}^{k+1}\ra^c_t=2d\xi_t\sum_{k=2}^{\infty}\frac{(i\lambda)^{(k-1)}}{(k-1)!} \la\hat{N}^{k}\ra^c_t=2d\xi_t\Big[\partial_{i\lambda}\mathcal{K}_t(\lambda)-\la\hat{N}\ra_t^c \Big]\label{eq:CGF_supp}
\end{align}
From the equation of the cumulant generating function, we obtain the evolution of the moment generating function, $\mathcal{Z}_t(\lambda)=e^{\mathcal{K}_t(\lambda)}$ as,
\begin{equation}            d\mathcal{Z}_t(\lambda)=2d\xi_t\Big[\partial_{i\lambda}\mathcal{Z}_t(\lambda)-\la\hat{N\ra_t\mathcal{Z}_t(\lambda)}\Big]. \label{eq:MGF_supp}
\end{equation}
We now perform the inverse Fourier transform of Eq.~\eqref{eq:MGF_supp} and integrating over $\lambda$ and obtain the evolution of $P(n,t)$ as,
\begin{align}
    dP(n,t)=2d\xi_t\big(n-\la\hat{N}\ra_t\big)P(n,t). \label{eq:Prob_dist_supp}
\end{align}

\section{Symmetry Restoration in other initial states}
\begin{figure}
    \centering
    \includegraphics[width=0.7\linewidth]{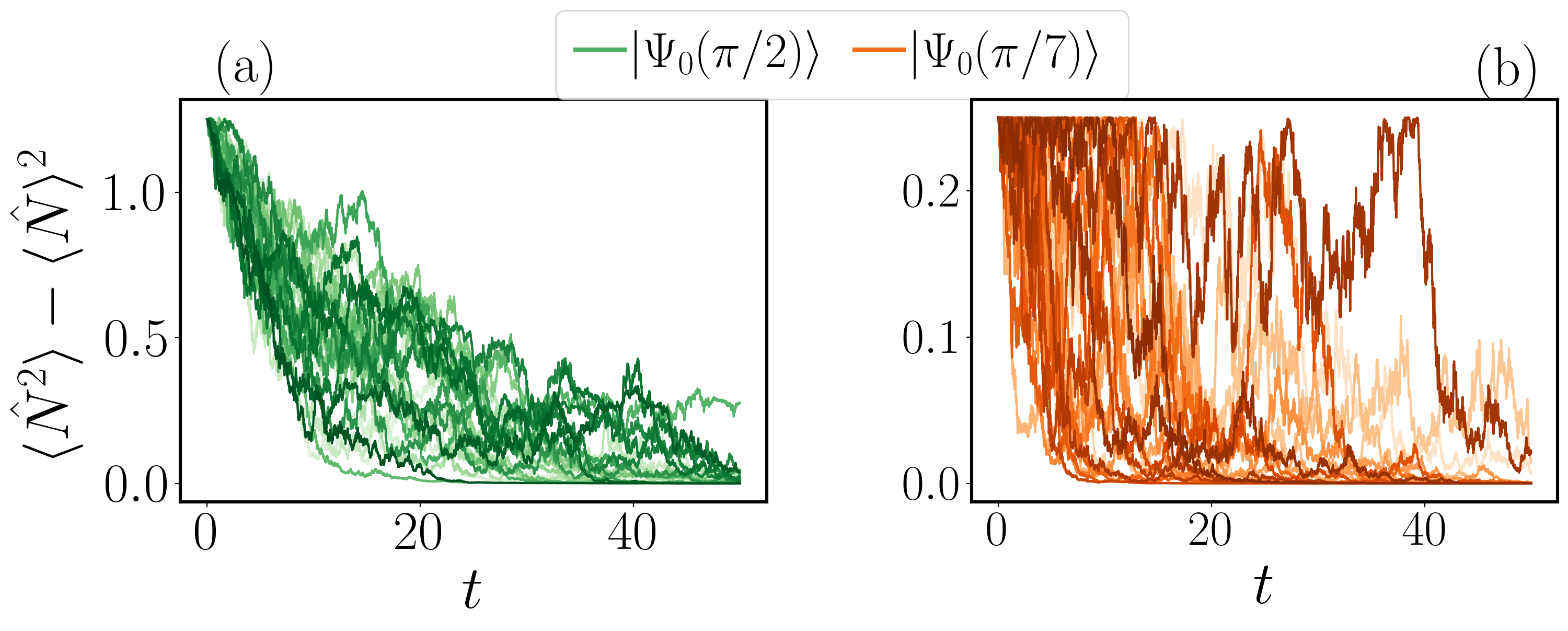}
    \caption{Number fluctuation is plotted with time $t$ for two different initial states, as defined in Eq.~\eqref{eq:ini_state_supp}, -- one is $|\Psi_0(\pi/2)\ra$ (green) which breaks the symmetry maximally and hence this state has high number fluctuation, the other state is $|\Psi_0(\pi/7)\ra$ which is less symmetry broken state with small number fluctuation. For both initial states, we have plotted $50$ different trajectories under global monitoring via QSD protocol. (a) Trajectories generated from the initial state $|\Psi_0(\pi/2)\rangle$ are plotted. (b) Trajectories generated from $|\Psi_0(\pi/7)\ra$ are plotted. We observe that although $|\Psi_0(\pi/2)\rangle$ has high number fluctuations, it relaxes in comparatively similar timescale that of $|\Psi_0(\pi/7)\rangle$. This is because both the initial states contain equidistant number sectors in the superposition. We have used $L=5$ and $\gamma=0.1J$ and performed exact diagonalization.}
    \label{fig:diff_init_supp}
\end{figure}

In this section, we present the results of the symmetry restoration dynamics under the QSD for two different initial states. The aim of this section is to show that the initial states with higher number fluctuations does not always show faster symmetry restoration. Here we consider the following initial state,
\begin{align}
    |\Psi_0(\theta)\ra=\Big(\cos\frac{\theta}{2}|0\ra+\sin\frac{\theta}{2}|1\ra\Big)^{\otimes L},\quad \theta\in [0,\pi/2]\label{eq:ini_state_supp}
\end{align}
For $\theta=0$, the initial state respects global $U(1)$ symmetry, whereas for $\theta=\pi/2$, the symmetry is maximally broken. Therefore, for any nonzero $\theta$, the initial state is $U(1)$ symmetry broken and contain all number sectors in the superposition. Hence the smallest separation between the number sectors is $n-m=1$ for all such states, leading to comparable time scales of symmetry restoration. We consider two different $\theta$ values and the corresponding initial states are, $|\Psi_0(\pi/2)\ra$ (highly symmetry broken state) and $|\Psi_0(\pi/7)\ra$ (less symmetry broken state). In Fig.~\ref{fig:diff_init_supp}, we observe that trajectories generated from $|\Psi_0(\pi/2)\ra$ and $|\Psi_0(\pi/7)\ra$ does not have significant difference in the symmetry restoration time-scale.

\section{Symmetry Restoration under the ensemble averaged dynamics}
In this section, we discuss the $U(1)$ symmetry restoration under the ensembled averaged Gorini- Kossakowski-Sudarshan- Lindblad (GKSL) quantum master equation. The GKSL equation that represents global dephasing by the jump operator $\hat{N}$ is given by,
\begin{equation}
    \partial\rho/\partial t =-i\big[\hat{H},\rho\big]+\gamma\,\Big[\hat{N}\rho  \hat{N}-\frac{1}{2}\{\hat{N}^2,\rho\}\Big], \label{eq:GKSL_global_supp}
\end{equation}
where $\gamma$ being the strength of the global dephasing. The global monitoring under QJ and QSD protocol and the global classical noise protocol [see End Matter] appear as different unravelings of the same GKSL equation, defined in Eq.~\eqref{eq:GKSL_global_supp}. One can, in principle, design an infinite number of different unravelings of the GKSL equation, each of which generates an ensemble of quantum trajectories. Thanks to the advancement of quantum technological platforms, implementation of different unravelings of the GKSL equation and realization of quantum trajectories have become possible~\cite{Rainer1986,Vijay2011,Kater2013}. 

\begin{figure*}
    \centering
    \includegraphics[width=0.25\linewidth]{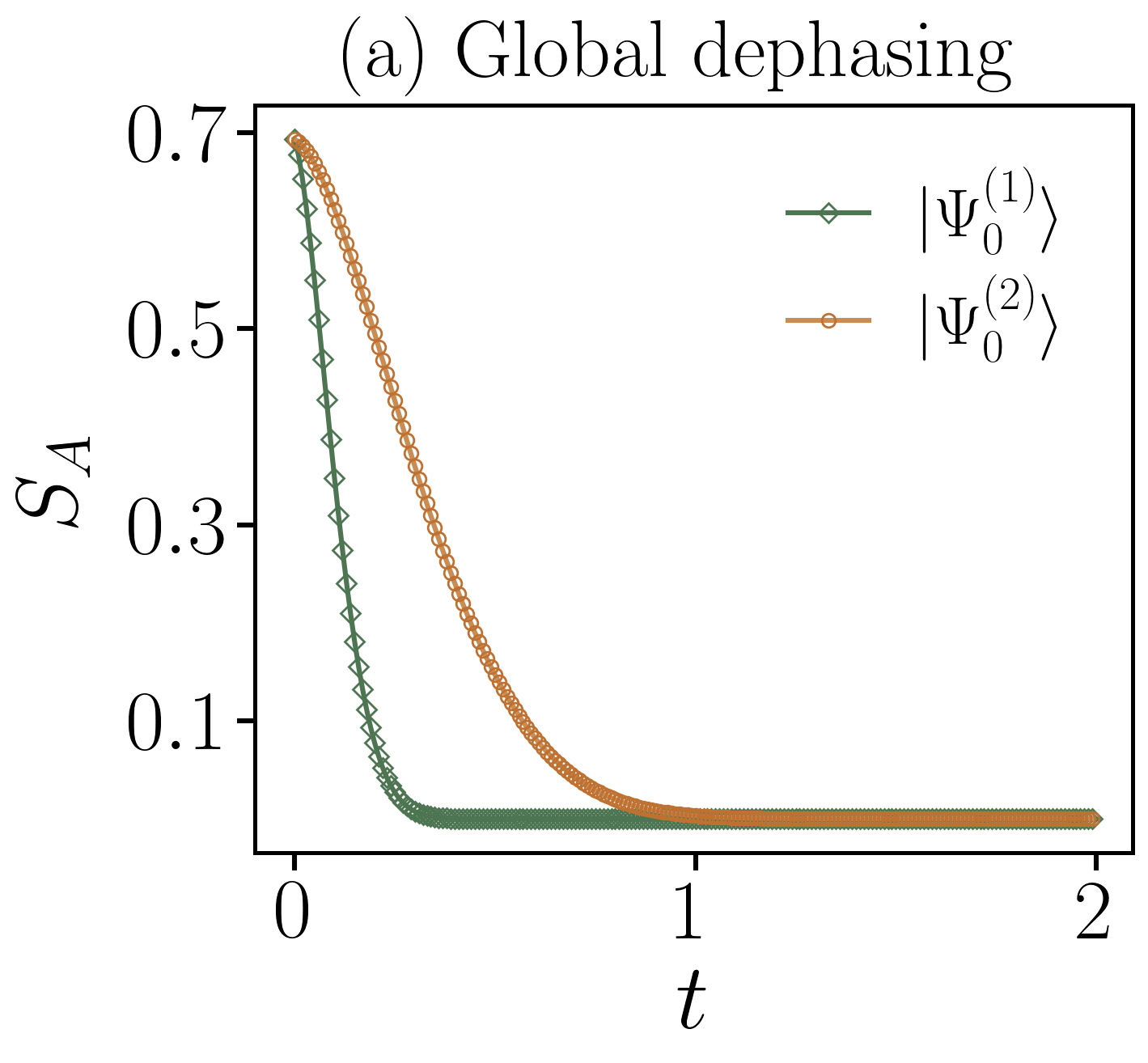}%
    \includegraphics[width=0.25\linewidth]{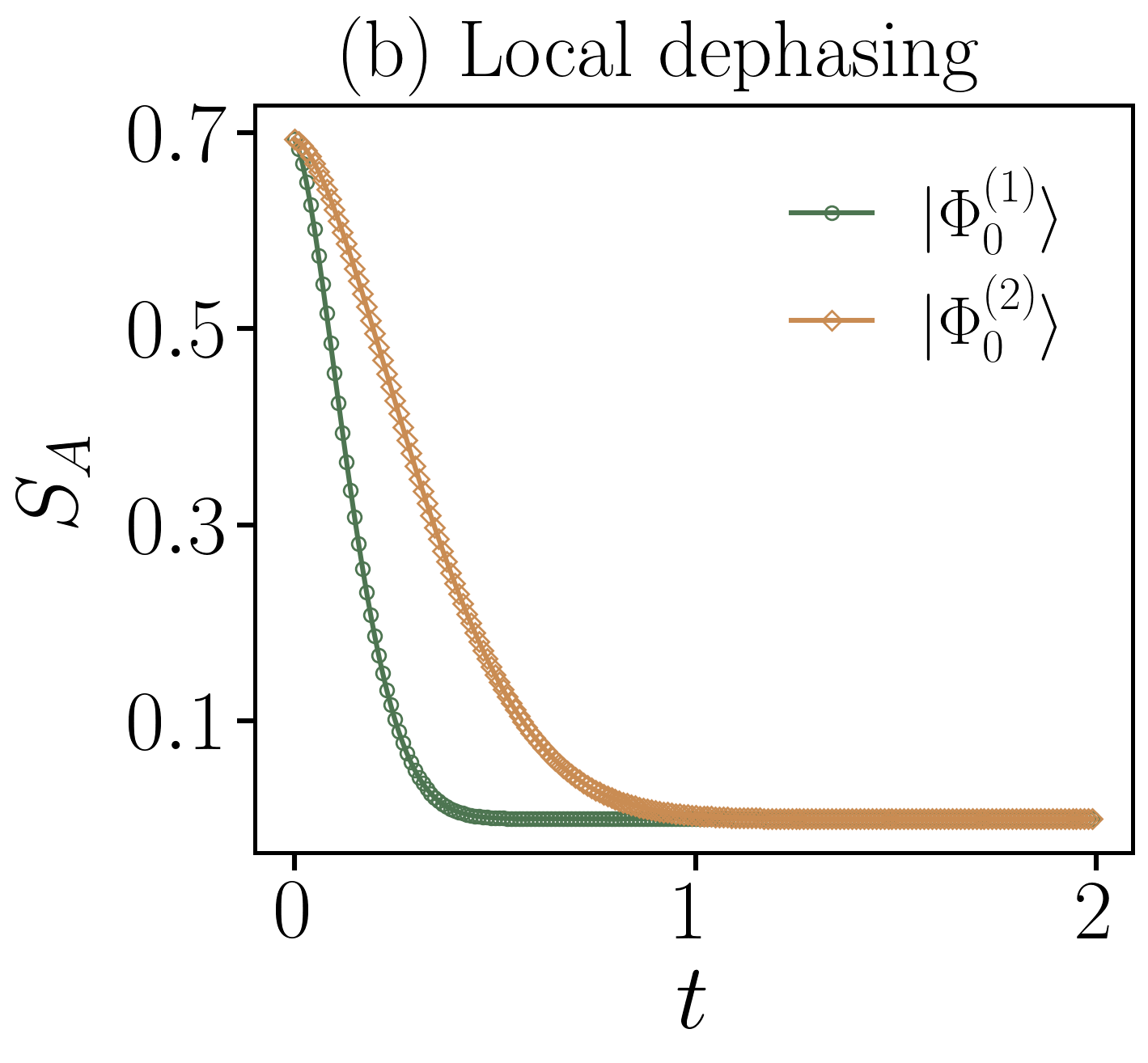}%
    \includegraphics[width=0.25\linewidth]{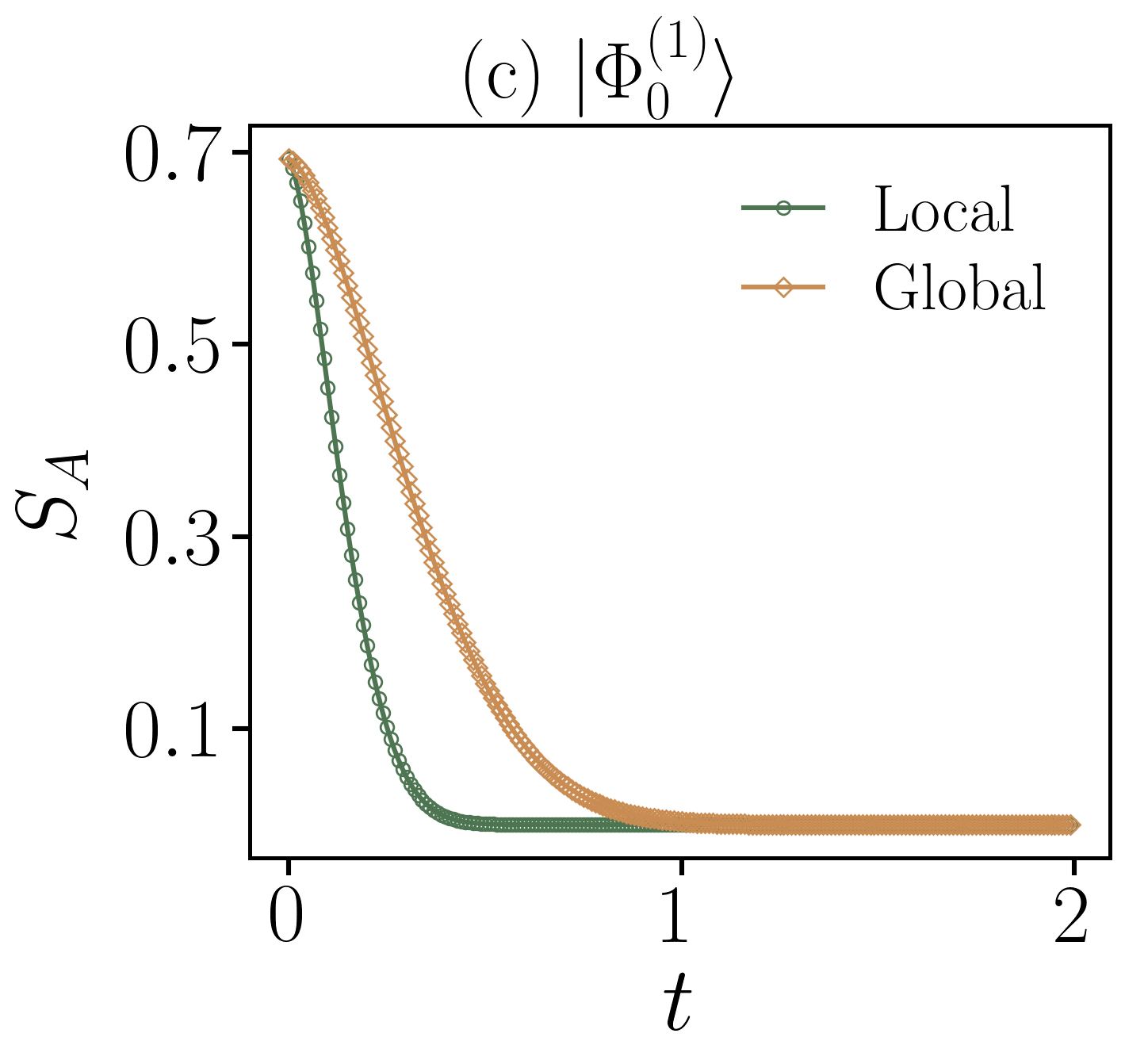}%
    \includegraphics[width=0.25\linewidth]{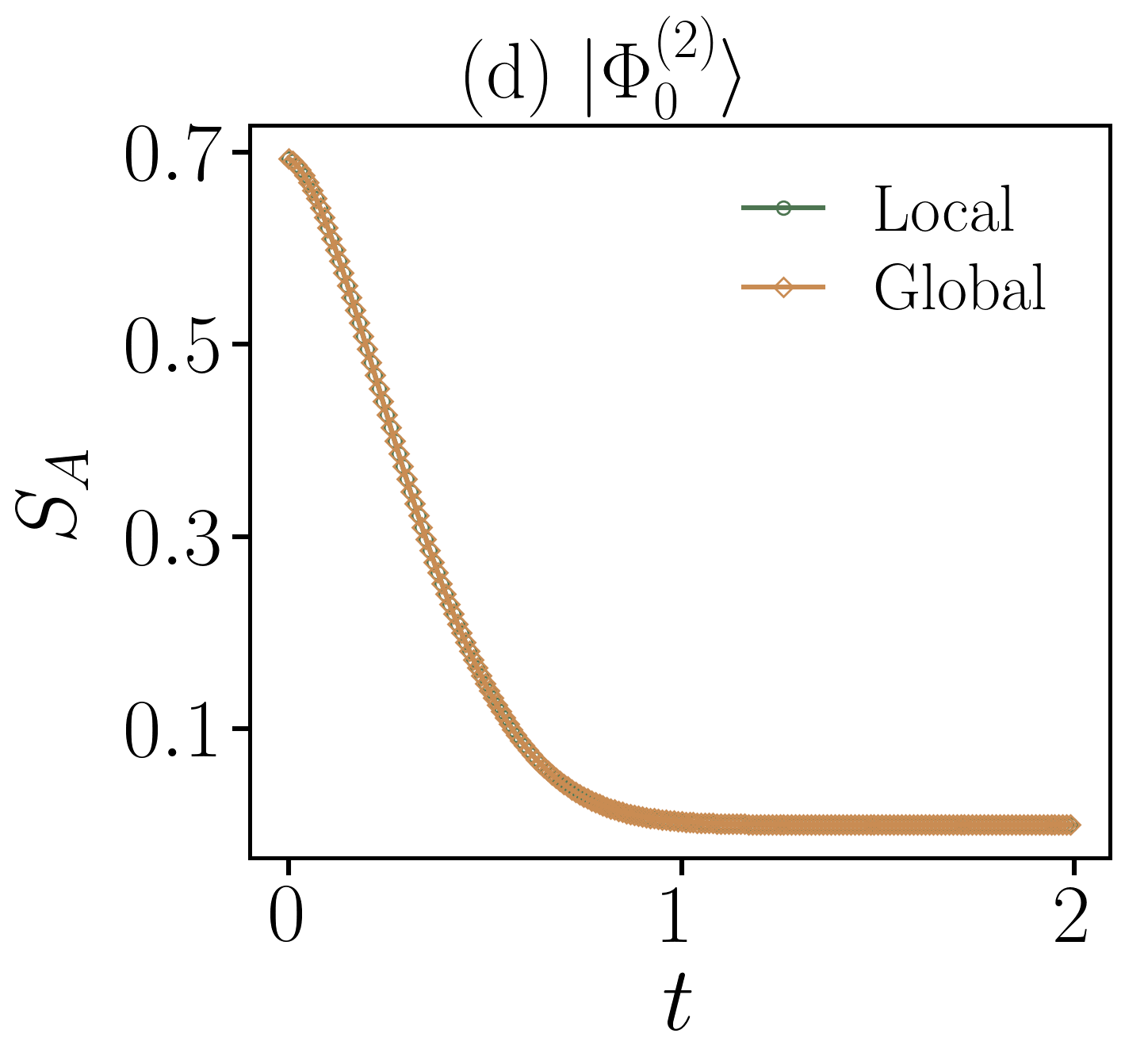}%
    \caption{Plot for the ensemble-averaged dynamics: The entropy asymmetry defined in Eq.~\eqref{eq:entropy_asym_supp} is plotted. (a) Under global dephasing defined in Eq.~\eqref{eq:GKSL_global_supp}, symmetry restoration dynamics is simulated for the initial states $|\Psi_0^{(1)}\ra$ and $|\Psi_0^{(2)}\ra$. (a) Global dephasing where the jump operator is $\hat{N}$. It shows that the initial state $|\Psi_0^{(1)}\ra$ relaxes faster than the initial state $|\Psi_0^{(2)}\ra$. (b) Local dephasing where the jump operators are $\hat{n}_i$ at every site $i$. It shows that the initial state $|\Phi_0^{(1)}\ra$ relaxes faster than the initial state $|\Phi_0^{(2)}\ra$ under local dephasing. (c) Comparison of local and global dephasing for the initial state $|\Phi_0^{(1)}\ra$ which shows that local dephasing is faster than global dephasing. (d) However, for the initial state $|\Phi_0^{(2)}\ra$, both global and local dephasing have similar symmetry restoration dynamics.}
    \label{fig:global_local_deph_supp}
\end{figure*}

We first discuss the case of global dephasing described by the GKSL equation in Eq.~\eqref{eq:GKSL_global_supp}.
Unlike QJ and QSD, under the GKSL equation, $d\la \hat{N}\ra/dt=0$. All moments and cumulants of $\hat{N}$ do not have any dynamics irrespective of the initial states. Hence, to observe symmetry restoration under the GKSL equation, one has to resort to the evolution of state $\rho(t)$ itself. 
Starting from a $U(1)$ symmetry broken initial state $|\Psi_0\ra=\sum_{n}c_n^0|n\ra$, the evolution of the population and coherences can be obtained from the Eq.~\eqref{eq:GKSL_global_supp} as,
\begin{align}
    &\frac{d}{dt}\la n|\rho|n\ra = 0,\quad\frac{d}{dt}\la n|\rho|m\ra = -\frac{\gamma}{2}(n-m)^2.\label{eq:coherence_evol_supp}
\end{align}
The absence of Hamiltonian in Eq.~\eqref{eq:coherence_evol_supp} is due to $U(1)$ conservation of the Hamiltonian. This confirms that under the averaged dynamics, a state containing superposition of distant number sectors always relax faster than a state containing superposition of nearby number sectors. Final steady state in such case is $\rho_{\rm ss}=\sum_n |\tilde{c}_n|^2|n\ra\la n|$. To numerically investigate this, we use the quantity entropy asymmetry~\cite{Calabrese2023}, which is defined as,
\begin{align}
S_A = -{\rm Tr}[\rho_{\rm sym}\ln \rho_{\rm sym}]+{\rm Tr}[\rho \ln \rho],\quad {\rm where}\quad\rho_{\rm sym}=\int_{-\infty}^{\infty} \frac{d\alpha}{2\pi}e^{-i\alpha \hat{N}}\,\rho\, e^{i\alpha\hat{N}}\label{eq:entropy_asym_supp}
\end{align}
In Fig.~\ref{fig:global_local_deph_supp}(a), we have plotted the entropy asymmetry $S_A$ for the the initial state $|\Psi_0^{(1)}\ra=\big(|1\ra+|L-1\ra\big)/\sqrt{2}$ which relaxes faster than the initial state $|\Psi_0^{(2)}\ra=\big[|(L-1)/2\ra+|(L+1)/2\ra\big]/\sqrt{2}$.

We next discuss the case of local dephasing in every site of the lattice. The GKSL equation describing this setup is given by,
\begin{align}
    \partial\rho/\partial t =-i\big[\hat{H},\rho\big]+\gamma\,\sum_{i=1}^{L}\Big[\hat{n}_i\rho  \hat{n}_i-\frac{1}{2}\{\hat{n}_i,\rho\}\Big]. \label{eq:GKSL_local_supp}
\end{align}
We once again consider the initial states $|\Phi_0^{(1)}\ra=\big[|\!\!\uparrow\downarrow\uparrow\downarrow\uparrow \ra+|\!\!\downarrow\uparrow\downarrow\uparrow\downarrow\ra\big)/\sqrt{2}$ and $|\Phi_0^{(2)}\ra=\big[|\!\!\uparrow\uparrow\uparrow\downarrow\downarrow \ra+|\!\!\uparrow\uparrow\uparrow\uparrow\downarrow\ra\big)/\sqrt{2}$ for the local dephasing case and investigate their symmetry restoration dynamics under the GKSL equation in Eq.~\eqref{eq:GKSL_local_supp}. Both of these states correspond to superposition of nearby number sectors but they have different overlap between the density profiles across the number sector. In Fig.~\ref{fig:global_local_deph_supp}(b), we observe that $|\Phi_0^{(1)}\ra$ restores the symmetry faster than $|\Phi_0^{(2)}\ra$. In Fig.~\ref{fig:global_local_deph_supp}(c) and (d), we compare the symmetry restoration under global and local dephasing. We observe that for the state $|\Phi_0^{(1)}\ra$, local monitoring can further accelerate the dynamics than global monitoring. However in $|\Phi_0^{(2)}\rangle$, both global and local monitoring leads to same dynamics of symmetry restoration.

\end{document}